Canadian Physics Counts:

Considering How Identity Relates to Experiences of Harm within the Canadian Physics Community

Authors

Adrianna Tassone[1], Eden J. Hennessey[1], Anastasia Smolina[2], Skye Hennessey[1], Kevin Hewitt[5], Shohini Ghose[3,4]

[1]Wilfrid Laurier University, Department of Psychology, Waterloo, ON, Canada, N2L 3C5

[2]University of Toronto, Department of Medical Biophysics, Toronto, ON, Canada, M5G 2C4

[3]Wilfrid Laurier University, Department of Physics and Computer Science, Waterloo, ON, Canada, N2L 3C5

[4]Quantum Algorithms Institute, Surrey, BC, Canada, V3T 5X3

[5]Dalhousie University, Department of Physics and Atmospheric Science, Halifax, NS, Canada, B3H 4R2

Corresponding Author: Eden Hennessey (ehennessey@wlu.ca)

**Abstract**

Harmful experiences such as harassment and discrimination continue to push many people out of science. To better understand identities and experiences of harm among physicists, we conducted Canadian Physics Counts: the first comprehensive national survey examining equity, diversity, and inclusion (EDI) within Canada's physics community. We explored experiences of harm focusing on personal harassment, sexual harassment, and sexual assault. We measured both direct experiences of harm and awareness of harm happening to others. Our analyses revealed that women and gender-diverse physicists reported experiencing personal harassment at twice the rate of men, a pattern consistent across all academic positions, including students and early-career researchers. An intersectional focus revealed even deeper inequities; Black women and men reported the highest rates of personal harassment, while Indigenous women and men faced elevated levels of sexual harassment. Physicists with disabilities were disproportionately affected: disabled women and gender-diverse respondents reported the highest rates of personal and sexual harassment and sexual assault, and disabled men experienced more personal harassment than men without disabilities. These findings are a clear call to action to the physics community to confront racism, sexism, homophobia, and ableism so every physicist can thrive and contribute to solving society's greatest challenges.

**Keywords:** *physics, diversity, inclusion, gender, race, disability, harassment, discrimination*

# Introduction

To advance scientific discovery and solve societal problems, the scientific community must attract and retain talented researchers. Equity, diversity, inclusion (EDI), and accessibility policies that enable all researchers to thrive are essential not only ethically but for scientific progress. This is why the Canadian federal government has committed to these principles through initiatives like the Dimensions Charter (Natural Sciences and Engineering Research Council, 2025a), the Canada Research Chairs EDI Action Plan (Canada Research Chairs, 2025) and the Natural Sciences and Engineering Research Council of Canada's Chairs for Inclusion (Natural Sciences and Engineering Research Council, 2025b). On an institutional level, to date, more than 60 post-secondary institutions across Canada have signed the Scarborough Charter to combat anti-Black racism (Scarborough Charter, 2025). Together, national and institutional initiatives and research evidence demonstrate the importance of attracting and retaining diverse talent. Team members with different backgrounds and perspectives can help groups solve problems, be more creative, and make better decisions (See Sulik et al., 2022 and van Knippenberg, 2024 for reviews). Team diversity can also lead to better, more original, and more influential scientific papers (Campbell et al., 2013; Hofstra et al., 2020; Yang et al., 2022). Additionally, the right to participate in science free from discrimination or harassment has been enshrined by the United Nations (United Nations, 2026).

Despite efforts to increase diversity and inclusion in science (see Meyer et al., 2025 and Palid et al., 2023 for reviews), there is still a lack of demographic diversity in physics in Canada (Aldosary et al., 2023; CAUT, 2018; Hennessey & Smolina et al., 2025; Perreault, 2018; Porter & Ivie, 2019; Strickland, 2017), and the United States (NCSES, 2023). While there are a multitude of contributing factors to attracting and retaining diverse scientists, differences in experiences can impact retention and there is ample evidence that experiences vary based on identity. Specifically, experiences of harm – the focus of the current paper – among underrepresented groups in physics (including those who are women, 2SLGBTQIA+, Black, Indigenous, or Persons of Colour [BIPOC], or disabled) are far too common.

For example, women in physics report a disproportionate number of experiences of harassment and discrimination. Aycock and colleagues (2019) surveyed undergraduate women in physics in the United States and found that 74% of respondents experienced at least one form of gender harassment in a physics education or workplace context. Paradis and colleagues (2023) interviewed medical physicists in the United States and found that women were more likely to report personal experiences of discrimination or harassment compared to men (75% and 44% respectively), as did Litzellachner and colleagues (2024) in their survey of STEM academics based in the United Kingdom. Further, women seeking advancement in their careers may be more likely to experience sexual harassment than other women (Raj et al., 2020), perhaps hindering their progression into leadership positions.

Racialized students in physics and biomedical sciences experience discrimination, racism, exclusion from crucial resources such as study groups, microaggressions, being undermined, and subtle discrimination such as receiving less time and attention from professors than other students (Park et al., 2020; Park et al., 2025; Rosa & Mensah, 2016). Black men studying science, technology, engineering, and math (STEM) report that the way they talk, walk, and dress in academic spaces is sometimes viewed with suspicion and fear, and that peers and instructors excessively question their work because they do not fit the stereotype of a scientist (Park et al., 2025; Spencer, 2021). Further, racialized physicists are more likely than White physicists to report hearing racist and homophobic comments (Clancy et al., 2017).

Queer physicists report exclusion, mockery, harassment, and being ignored (Barthelemy, 2020; Barthelemy et al., 2022, Gutzwa et al., 2024). Barthelemy and colleagues (2022) found that 22% of 2SLGBTQIA+ physicists in their research had experienced exclusionary behaviours, and this was particularly common among women and gender non-conforming respondents. Disturbingly, this number more than doubled for transgender respondents (49%). The structure of work and study in physics as well as attitudes and interpersonal interactions are examples of ableism that excludes disabled physicists (Jeannis et al., 2020), such as reports of professors who make discriminatory comments about students who request accommodations (Guthrie et al., 2025). Unfortunately, the evidence is clear that harmful experiences including discrimination, harassment, and bullying remain widespread in physics and other scientific settings, underscoring the urgent need for meaningful change.

Harmful experiences in educational and workplace settings are linked to leaving physics (Park et al., 2020) and other STEM fields (Litzellachner et al., 2024; Minnotte & Pedersen, 2023). This is unsurprising given such experiences result in lower confidence in ability to graduate among college students (Jackson et al., 2023), increased mental and physical health problems among STEM faculty (Corbett et al., 2024; Minnotte & Pedersen, 2023), and higher levels of imposter feelings and reduced sense of belonging among undergraduate women in physics (Aycock et al., 2019). Conversely, sense of belonging is tied to academic outcomes in physics (Feser & Bauer, 2025; Seyranian et al., 2018), and intentions to remain in STEM (Litzellachner et al., 2024, Mattheis et al., 2022). Research on sexual harassment in biology and ecology shows links to lower professional engagement and delays in completing graduate programs (Wilkins et al., 2023), as well as decreased STEM motivation and career aspirations and trajectory (Leaper & Starr, 2019; Wilkins et al., 2023). Studies in non-STEM workplaces likewise connect harassment to depression, sleep problems, musculoskeletal injuries (Gale et al., 2019), psychological distress, and alcohol misuse (Rospenda et al., 2023). These findings highlight the serious personal and professional consequences of harmful experiences, emphasising the need for effective prevention and remedial measures.

Previous research has long provided compelling evidence that those with multiple marginalized identities are at higher risk of harmful experiences in science, exemplifying intersectionality theory (Allen et al., 2022a; Charleston et al., 2014; Crenshaw, 1991; Malcom & Malcom, 2011; Ong et al., 2011; Ovink et al., 2024; Wilkins-Yel et al., 2019). Clancy and colleagues (2017) surveyed members of the American Physical Society and found that racialized women were the most likely to report verbal harassment related to their race and gender, compared to all other groups. They also found that racialized women were the most likely to report feeling unsafe in their workplaces due to their gender or race. Women of all races, but particularly racialized women, reported skipping professional events including meetings, fieldwork, or classes because they did not feel safe attending, which can have negative consequences for professional advancement.

In other examples, Allen and colleagues (2022a) found that Black women in four-year academic institutions in the United States experienced both racism and sexism from professors and peers. One qualitative meta-synthesis covering 42 studies found that BIPOC women in academia faced marked barriers including limited access to identity-congruent mentors and role models, funding, exposure to tokenism, microaggressions, high teaching and service loads, and bias in promotion and tenure criteria (Corneille et al., 2019). Examining these experiences, Wilkins-Yel et al., (2019) found that racialized women in STEM faced both hyper *invisibility* (being ignored or excluded) and hyper *visibility* (being scrutinized or tokenized), creating a unique and fraught double-bind. The authors interpreted their findings through intersectional invisibility theory (Purdie-Vaughns & Eibach, 2008), which suggests that BIPOC women are less likely to fit the typical image of a scientist in terms of both race and gender, making them more susceptible to discrimination. In science, being perceived as competent is essential, however, some research has found that Black women were perceived as less warm and competent compared to their peers (Eom et al., 2025). In contrast, Cech (2022) used survey data from STEM professionals based in the United States and found that White, heterosexual men without disabilities were most likely to experience social inclusion, professional respect, and access to career opportunities, compared to other groups.

Similar to the intersections of race and gender, research on disability in the workplace (outside of physics) has found that women with disabilities report lower pay, more workplace stress, and less autonomous working conditions compared to men with disabilities (Brown & Moloney, 2019). Overall, past research makes it clear that understanding experiences of harm with an intersectional focus is crucial to more fully understanding the physics landscape in Canada and potential interventions to support equity, diversity, inclusion, and accessibility.

While personally experiencing discrimination or harassment is obviously harmful, witnessing harm can also have negative psychological and health impacts. Acquadro Maran and colleagues (2022) found that witnessing workplace sexual harassment was associated with negative outcomes including burnout, lower life satisfaction, and negative emotions. Similarly, a recent meta-analysis (Nielsen et al., 2024) found that witnessing workplace bullying was associated with mental health and somatic problems, sleep disturbances, and lower well-being. Considering that witnessing and experiencing discrimination both have harmful effects, it is important to assess the extent to which both types of experiences occur in science, including in physics. We refer to witnessing harm, hearing a disclosure of harm from the involved persons, or hearing about harm another way (e.g., through a third-party) as 'awareness of harm.' We conceptualized awareness of harm as a climate-relevant indicator that captures informal knowledge networks within institutions, sometimes referred to as "whisper networks" (Johnson, 2023; Jung & Mendoza 2023).

Previous research has provided valuable insights into physicists' experiences, but important gaps remain. Much of the existing research has been conducted in the United States, where demographics and institutional contexts differ from Canada. One exception, Aldosary and colleagues (2023), reported that 17% of Canadian medical physicists experienced or were aware of sexual harassment, but small samples limited intersectional analysis. Robust empirical data from the Canadian context are paramount to developing policies and programming that effectively reduce harmful experiences and increase inclusion. Thus, a priority of the current research was to document experiences and awareness of harm in Canada using an intersectional approach so that this information can guide how we prioritize and collectively respond to harm in science. To this end, in the current study, we measured physicists' personal experiences of harm and awareness of harm in their current place of work or study including personal harassment, sexual harassment, and sexual assault, representing a broad spectrum of discrimination. These outcomes were chosen because sexual harassment is a persistent problem in physics environments (Aycock et al., 2019; Aldosary et al., 2023; Clancy et al., 2017; Barthelemy, 2020; Barthelemy et al., 2022; Paradis et al., 2023) and other workplaces (Statistics Canada, 2024), affecting predominantly women (Statistics Canada, 2024), with negative impacts on career progression and retention (e.g., Litzellachner et al., 2024; Leaper & Starr, 2019; Wilkins et al., 2023). As such, sexual harassment was a priority to assess in the Canadian context. We also measured experiences of personal harassment, given that discriminatory events can also occur based on race, disability, sexual orientation, and other aspects of identity.

Although there are exceptions from outside of Canada and from different fields (e.g., Cech, 2022; Clancy, 2017; Wilkins et al., 2023) previous work conducted on Canadian physicists has not often taken an intersectional approach (e.g., Aldosary et al., 2023). While still informative, such inquiries do not allow for assessing the interactive effects of multiple underrepresented identities, which is essential given that discrimination may occur based on intersecting demographics (e.g., race and gender, gender and disability). In the current research, we analyze responses from 1968 physicists across Canada who participated in a larger demographic survey (see Hennessey & Smolina et al., 2025), allowing for some intersectional statistical comparisons. Although there are endless intersections that comprise people's unique identities and impact their experiences (e.g., a Black, disabled, man; a non-binary, queer, racialized person), we chose to focus on gender and include intersections of race, disability, and sexuality (i.e., comparing BIPOC women, BIPOC men, White women, White men; disabled women, disabled men, and men and women without disabilities). We did so due to sample size limitations when considering other intersections, and the consistent statistical effect of gender.

Guided by previous research (Allen et al., 2022a; Cech, 2022; Clancy et al., 2017; Eom et al., 2025; Leaper & Starr, 2019; Park et al., 2020; Wilkins et al., 2023) and intersectional theory (Crenshaw, 1991), we expected that respondents with multiple underrepresented identities (e.g., women, gender-diverse, BIPOC, disabled, or sexually diverse and their intersections) would report experiencing harm more frequently than those with fewer, or no, underrepresented identities.

## Method

### Procedure

The current study draws on data from a larger survey of Canadian physicists. Demographics were reported in detail in a previous publication (Hennessey & Smolina et al., 2025). Here, we focus specifically on physicists' experiences and awareness of harassment and assault, referred to collectively as harm. The broader survey, approved by the Dalhousie University Research Ethics Board (20205261), was distributed by email to members of the Canadian Association of Physicists (CAP), all Canadian physics departments, institutional partners (e.g., Perimeter Institute, SNOLAB, TRIUMF, Canadian Light Source, Institute for Quantum Computing, Stewart Blusson Quantum Matter Institute), CAP's network of high school physics teachers, and professional associations (e.g., Canadian Organization of Medical Physicists, Canadian Astronomical Society). Eligible participants either held a physics degree and lived or worked in Canada or were pursuing a physics degree at a Canadian institution. After providing informed consent, participants completed measures assessing their demographics and experiences of harm within the Canadian physics community. To increase survey response rates, participants were encouraged to share the survey link within their professional networks. The survey was available in French and English, with French entries translated into English for analysis.

**Participants**

Given the sensitive nature of the subject matter, respondents could skip any questions. Out of the 2532 people who completed larger survey, 1968 (77.7%) answered the harm-related questions. Table 1 compares the full sample with this subsample.

To assess whether identity predicted skipping the harm questions, we conducted four Chi-square tests of independence, comparing observed and expected frequencies. Gender was significantly associated with skipping, $X^2$ (2, $N = 2491$) = 23.9, $p < .001$, $\varphi = .10$: men were overrepresented among those who skipped (386 vs. 337 expected), whereas women (149 vs. 194) and gender diverse participants (15 vs. 19 expected) were underrepresented. Race was also associated with skipping, $X^2$ (1, $N = 2484$) = 73.8, $p < .001$, $\varphi = .17$: BIPOC participants were overrepresented among those who skipped (254 vs. 171 expected), and White participants were underrepresented (290 vs. 372 expected). Disability was associated with skipping, $X^2$ (1, $N = 2265$) = 5.6, $p = .02$, $\varphi = .05$: disabled participants were underrepresented among those who skipped (26 vs. 39 expected) and participants without disabilities were overrepresented (473 vs. 460 expected). Sexual orientation was associated with skipping, $X^2$ (1, $N = 2449$) = 16.4, $p < .001$, $\varphi = .08$: sexually diverse participants were underrepresented among those who skipped (76 vs. 109 expected), and heterosexual participants were overrepresented (462 vs. 429 expected).

Some respondents may have skipped the harm questions because they felt it was irrelevant or because they preferred not to disclose harmful experiences. Notably, most underrepresented groups (women, gender diverse, disabled, and sexually diverse respondents) were less likely to skip, whereas BIPOC respondents were more likely to skip. Because motivations for skipping cannot be determined and effect sizes were small, we interpret these findings with caution. The demographic descriptions below reflect the 1968 participants who completed the harm questions.

*Age*

Age was reported in brackets from under 20 to over 70. About half (49.8%) of respondents were 20-29 years old, consistent with the large proportion of students (60.7%).

*Gender*

Respondents identified as men (57.9%), women (37.1%), gender diverse (3.7%; e.g., non-binary, genderqueer, agender), preferred not to answer (1.2%) or did not respond (.2%). Although men were more likely to skip the harm questions, they remained the largest group.

*Race/Ethnicity and Generation Status*

Respondents were White (71.6%), South Asian (7.3%), East Asian (4.9%), Latin American (2.1%), West Asian (1.6%), Arab (1.5%), Southeast Asian (.7%), Black (1.1%), Indigenous (.3%), 6.4% selected multiple categories, 1.2% preferred not to answer, .1% did not respond, and .9% preferred to self-describe. Regarding generation status, 30.8% reported both Canadian parents and grandparents, 15% Canadian parents only, 9.4% one Canadian parent, 8.9% non-Canadian parents, .1% other, and 35.9% did not answer.

*Disability*

Most respondents (82.1%) reported no disability, 7.6% reported a disability, 3.4% were unsure, 6% did not answer, and .9% preferred not to answer.

*Sexual Orientation*

Respondents identified as heterosexual (76.3%), bisexual (7.2%), gay (2.0%), lesbian (1.3%), pansexual (1.1%), asexual (1.1%), queer (.8%), questioning (.8%), and polyamorous (.1%). Some selected multiple responses (7.1%), preferred not to answer (1.8%), or did not respond (.4%).

**Table 1**

*Comparison of Race/Gender Demographics of the Total Sample and Subsample of Those Who Responded to the Harm Questions*

|  | Larger Survey Sample | | Subset who Responded to the Harm Questions | |
|---|---|---|---|---|
|  | N | % | N | % |
| **BIPOC Gender Diverse** | 23 | .9% | 18 | .9% |
| **BIPOC Women** | 290 | 11.8% | 213 | 11.1% |
| **BIPOC Men** | 463 | 18.8% | 294 | 15.3% |
| **White Gender Diverse** | 62 | 2.5% | 53 | 2.8% |
| **White Women** | 578 | 23.5% | 508 | 26.4% |
| **White Men** | 1045 | 42.5% | 839 | 43.6% |
| **Total** | 2532 | 100% | 1925 | 100% |

*Notes*. Participants who did not report their race or gender are not included in this table. Totals may not equal exactly 100% due to rounding.

## Measures

Those who completed the harm questions were shown definitions from the Dalhousie University Sexualized Violence Policy and Personal Harassment policies covering personal harassment (e.g., bullying, intimidation), sexual harassment (e.g., unwanted sexual advances), and sexual assault (e.g., unwelcome sexual contact; full definitions in the Supplementary Materials). For each type of harm, one item assessed frequency of personal experiences (i.e., I have experienced personal harassment/sexual harassment/sexual assault in my place of work/study), and another assessed awareness of others' experiences (i.e., I know somebody else has experienced personal harassment/sexual harassment/sexual assault in my place of work/study). Participants rated each statement on a scale from 1 = *never*, 2 = *once*, 3 = *occasionally*, 4 = *regularly*, 5 = *frequently*. These items were modified from Cech and Waidzunas (2019). An optional comments box followed.

### Data Analysis Approach

#### Demographic Groupings

Some demographic variables (gender identity, race, disability, sexual orientation) were aggregated to protect anonymity and increase statistical power, such as grouping racialized identities into a BIPOC (i.e., Black, Indigenous, and People of Colour) category (e.g., Centre for Research and Innovation Support, 2024; Clancy et al., 2017; Coleman & Yantis, 2024; Jin et al., 2024; Levandowski et al., 2024; Magoon et al., 2022). We acknowledge that this practice is imperfect, as collapsing diverse identities into a single category can inadvertently reinforce Whiteness as the normative standard and position all other identities as a monolithic "other," obscuring the distinct experiences of different racialized groups in STEM. For example, students report varying types and frequencies of microaggressions by racial identity (Lee et al., 2020). Nevertheless, we assert that comparisons between White and BIPOC respondents are still valuable, as BIPOC communities may experience shared underrepresentation and discrimination not typically experienced by non-racialized people (e.g., American Institute of Physics, 2020; Malcom & Malcom, 2011; Ong et al., 2011). Thus, we sometimes combined racial identity groups, while recognizing that fully disaggregated racial data would ultimately provide more meaningful insights.

Similarly, we employed the term 'gender diverse' to describe those with gender identities distinct from 'man' or 'woman' (i.e., non-binary, genderqueer, agender, etc.), as gender diverse people report more harm in daily life than cisgender people on average (Prokopenko & Hango, 2022; Jaffray, 2020), making it important to assess their experiences in physics. However, small sample sizes limited intersectional analyses. Disability and sexual orientation were also grouped into binary categories described below.

### *Quantitative Data Analysis*

Analyses were conducted in IBM SPSS 29. Gender identity was examined using three categories (women, gender diverse, men), or two categories (women, man) across variables, including academic position, race (BIPOC and White), disability (disabled or not), and sexual orientation (sexually diverse and heterosexual). For each harm type, we report percentages of respondents experiencing or being aware of at least one instance. We combined these into a single category for simplicity and because even one instance of harm can have serious implications for mental and physical health, retention in the field, and sense of safety and comfort in the workplace. In other words, frequencies represent the percentage of each demographic group who have had a particular experience at least once. The means and standard deviations represent the average frequency with which each demographic group has had a particular experience, on the scale from never to frequently.

Intersectional analyses used four-group comparisons (e.g., BIPOC women, BIPOC men, White women, White men). Due to violations of statistical assumptions and unequal sample sizes, Kruskal-Wallis (KW) tests with Dunn post-hoc comparisons and Bonferroni corrections were applied. Academic positions and gender were analysed with Mann-Whitney U tests. We used $p < .05$ as the significance threshold, noting that Bonferroni corrections may be overly conservative in some cases; adjusted and unadjusted p-values are reported where appropriate. The KW test is a non-parametric statistical test used to determine whether there are statistically significant differences between the distributions of three or more independent groups. The test ranks all values in the data set (ignoring group membership; i.e., BIPOC woman; White man) and then compares the mean ranks for each group. The test produces an H-statistic, which is compared to the Chi-square distribution to assess statistical significance. Power analyses using G*Power software indicated that 1721 participants were required to detect small effects. Although our sample exceeded this at $N = 1968$, demographic imbalances may reduce power.

Our analyses addressed five central research questions:

1. How did frequencies of experiences and awareness of harm differ across gender identities?

2. How did frequencies of experiences and awareness of harm differ across binary gender identities and academic position?

3. How did frequencies of experiences and awareness of harm differ across binary gender identities and race?

4. How did frequencies of experiences and awareness of harm differ across binary gender identities and disability?

5. How did frequencies of experiences and awareness of harm differ across binary gender identities and sexual orientation?

### *Qualitative Data Analysis*

Respondents could provide comments in a text-box that did not provide a specific prompt to be non-prescriptive. Only 9% (*n* = 177) provided open-ended comments. Due to the small number of responses and absence of a specific prompt, we did not conduct a formal thematic analysis and instead use comments illustratively throughout the results. Some comments were edited for clarity and to remove any potentially identifying information. Only comments from respondents who agreed to have their quotes published are included.

## Results

The detailed statistical results of all analyses are in Supplementary Materials. A summary of findings is reported here. As a reminder, we present both percentages and descriptive statistics. For each type of harm, percentages represent the proportion of respondents in each demographic group who have personally experienced or been made aware of an instance of harm at least once. Means and standard deviations represent the average frequency with which each demographic group personally experienced a type of harm or was aware of an instance of harm, on a scale from (1) *never* to (5) *frequently*. Results are organized according to the five primary research questions.

### 1. How Did Frequencies of Experienced Harm and Awareness of Harm Differ Across Gender Identities?

Notably, there were no groups that did not experience harm. Women and gender diverse people reported a disproportionate amount of harm; about 2x the proportion of women and gender diverse respondents reported experiencing personal harassment compared to men. About 6x the proportion of women and 5x the proportion of gender diverse respondents compared to men reported sexual harassment, and 5x the proportion of women and 3x the proportion of gender diverse respondents compared to men reported sexual assault in their place of work or study.

Pairwise comparisons showed that women and gender-diverse people experienced personal harassment and sexual harassment with similar frequency to each other ($ps > .05$), and significantly more frequently than men ($ps < .01$). Women experienced sexual assault more frequently than men ($p < .001$) and were not different from gender diverse respondents ($p = 1.00$). Gender diverse respondents and men did not significantly differ ($p = .97$). Further, compared to men, higher proportions of women and gender diverse respondents reported experiencing all types of harm at least once or more (see Table 2 and Figure 1).

Women and gender diverse respondents reported more awareness of personal harassment and sexual harassment than men ($ps <.001$), with gender diverse respondents reporting more awareness of personal ($p = .04$) and sexual harassment ($p = .006$) than women. In other words, gender diverse respondents reported being aware of personal and sexual harassment most frequently of all gender groups. Compared to men, higher proportions of women and gender diverse respondents reported being aware of personal and sexual harassment at least once or more (see Table 2 and Figure 1). Gender diverse respondents were most frequently aware of sexual assault – significantly more so than both women and men ($ps < .001$), who did not differ from one another ($p = 1.00$). Compared to men and women, a higher proportion of gender diverse respondents reported being aware of sexual assault at least once (see Table 2 and Figure 1). See Table 2 for mean frequencies and standard deviations and Supplementary Materials for detailed statistical results of the Kruskal-Wallis tests.

**Table 2**

*Descriptive Statistics and Frequencies for Type of Harm for the Total Sample and by Gender Identity*

| | Total Sample (*N* = 1968) | | Men (*n* = 1139) | | Women (*n* = 730) | | Gender Diverse (*n* = 72) | |
|---|---|---|---|---|---|---|---|---|
| Outcome | *M* | *SD* | *M* | *SD* | *M* | *SD* | *M* | *SD* |
| **Experienced Personal Harassment** | 1.44 | .85 | 1.27 | .67 | 1.68 | 1.01 | 1.70 | 1.01 |
| *Never* | 74.3% | | 83.1% | | 62.1% | | 60.0% | |
| *Once or more* | 25.7% | | 16.9% | | 37.9% | | 40% | |
| **Awareness of Personal Harassment** | 1.90 | 1.05 | 1.79 | .98 | 2.03 | 1.11 | 2.37 | 1.18 |
| *Never* | 49.8% | | 53.3% | | 46.2% | | 32.9% | |
| *Once or more* | 50.2% | | 47.7% | | 53.8% | | 67.1% | |
| **Experienced Sexual Harassment** | 1.16 | .55 | 1.04 | .26 | 1.34 | .78 | 1.26 | .63 |
| *Never* | 90.1% | | 96.8% | | 80.4% | | 84.3% | |
| *Once or more* | 9.9% | | 3.2% | | 19.6% | | 15.7% | |
| **Awareness of Sexual Harassment** | 1.49 | .85 | 1.40 | .75 | 1.57 | .93 | 1.96 | 1.16 |
| *Never* | 70.6% | | 73.9% | | 67.4% | | 51.4% | |
| *Once or more* | 29.4% | | 26.1% | | 32.6% | | 48.6% | |
| **Experienced Sexual Assault** | 1.04 | .27 | 1.01 | .12 | 1.08 | .41 | 1.04 | .27 |
| *Never* | 97.7% | | 99.1% | | 95.4% | | 97.1% | |
| *Once or more* | 2.3% | | .9% | | 4.6% | | 2.9% | |
| **Awareness of Sexual Assault** | 1.17 | .57 | 1.15 | .51 | 1.19 | .59 | 1.51 | .98 |
| *Never* | 89.4% | | 90.4% | | 89.3% | | 73.9% | |
| *Once or more* | 10.6% | | 9.6% | | 10.7% | | 26.1% | |

*Note.* Sample sizes (n's) may vary as some participants skipped certain questions.
Some participants included in the 'total sample' did not report their gender, so the total *n* reported (*n* = 1968) is higher than the three gender groups combined (*n* = 1941).
The total of the three gender groups in Table 2 (*n* = 1941) is higher than in Table 1 (*n* = 1925) because some respondents reported gender but not racial identity.

**Figure 1**

*Percentage of Respondents Reporting One or More Instances of Harm by Gender and Type of Harm*

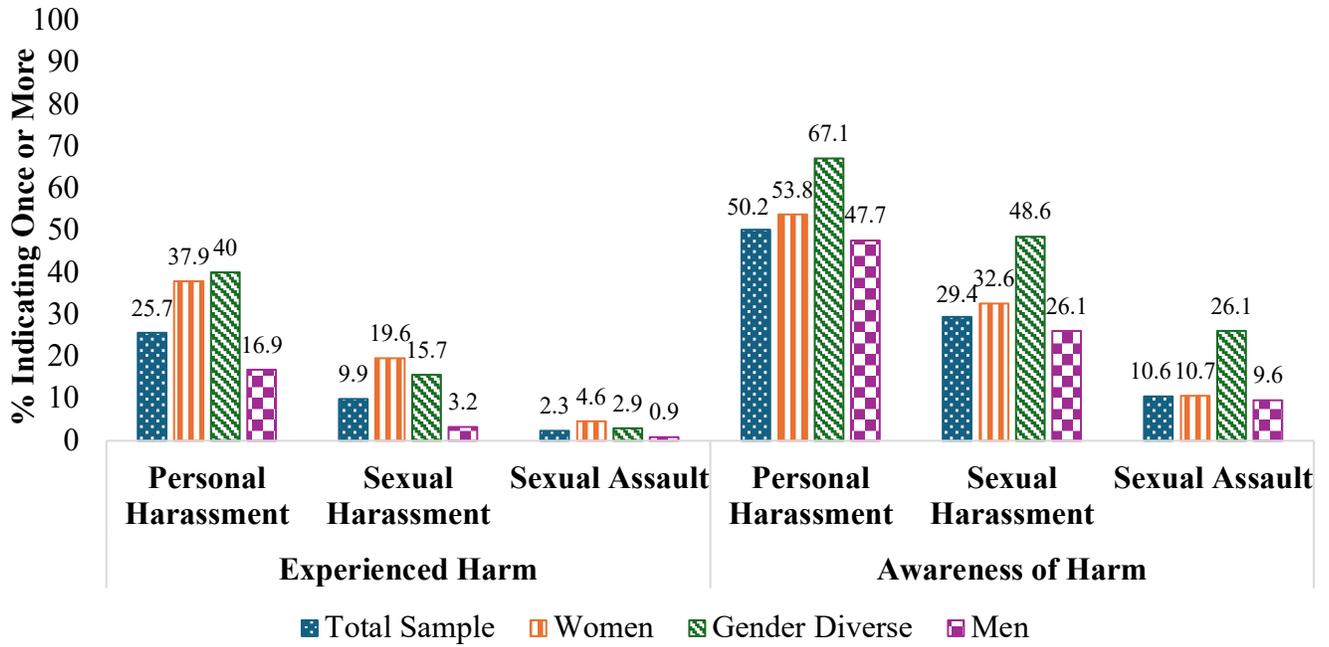

**Table 3**

*Qualitative Comments Illustrative of Experienced Harm*

| Comment | Participant Number & Gender Identity |
| --- | --- |
| When I was working on my Master's degree, a PhD student in my lab asked me to 'make a baby' with him. This influenced my decisions to not [pursue] a PhD. | Participant 1640, woman |
| One of my colleagues has taken it upon herself to spread/propagate mistruths about my teaching and other aspects of my interaction with students to others in my department.  Being a man, this was and is difficult to defend without coming across as being sexist; indeed I was accused of that when I displayed publicly -and admittedly, unprofessionally my anger and frustration by her lies. This experience has been both intimidating and belittling | Participant 581, man |
| Most memorable moment — feeling pressured to publicly kiss and feeling forced to date a teaching assistant to gain fair marks. I felt that if I didn't continue to engage with this person that he wouldn't mark me fairly. I didn't feel like the professor of the course would understand or know what to do. I stopped engaging with this TA and distancing myself from the course. I did pass but it is one of my worst marks from that degree. | Participant 1854, woman |
| Overheard [a] male student explaining to other male students why women should not be scientists. I was in the room and am a woman [....] the other men were very apologetic. Frequent harassment because my supervisor … uses his power over me regularly - e.g., yelling at me for not doing things fast enough or the way he would have done it, accusing me of lying on job application because of the way he interpreted it, promising collaborators immediate work from me on projects weeks before my thesis is due. | Participant 1665, woman |
| I have been a target of department-wide bullying for several years. The people who started the bullying were students and post-docs of senior professors, so my supervisor told me that I cannot speak to anyone about the issue. My health and productivity have suffered tremendously as a result. | Participant 1866, woman |

**Table 4**

*Qualitative Comments Illustrative of Awareness of Harm*

| Comment | Participant Number & Gender Identity |
|---|---|
| We work as TA's. Funnily enough I and a fair few other female grad students have experienced sexual harassment from students that we teach. | Participant 1656, woman |
| Basically, every woman I know has experienced such treatment. | Participant 2515, non-binary |
| I am aware of two female students who were sexually assaulted by another student in our program at off-campus social events. | Participant 2222, woman |
| It was a common in my undergraduate years to see fellow female undergrads getting their opinions or views or inputs dismissed because they were "just a girl". A few of them have confided in me that this was a big part in why they were choosing not to continue on in academia. | Participant 229, woman |
| At my current industry job, I have witnessed gender discrimination in the form of more senior personnel expressing outdated/stereotypical views of females -I am male, and have spoken to my female colleagues about it to see what their reaction was -they found it to be discriminatory. When I was in astronomy academia I heard *many* first- or second-hand accounts of sexual harassment of female colleagues taking place, but never witnessed it directly. | Participant 961, man |

## 2. How Did Frequencies of Experienced Harm and Awareness of Harm Differ Across Binary Gender Identities and Academic Position?

Women reported more frequent personal harassment experiences than men across all seven tested academic positions, and more frequent sexual harassment than men across most academic positions (all $ps < .05$, the exception being assistant professors among whom there was no gender difference; see Table S1 in Supplementary Materials). Women BSc and MSc students experienced a higher frequency of sexual assault than their men peers ($ps < .05$). At no level of academia did men report more harm than women, although importantly, men still experienced harm too — just not as frequently as women. See Figures 2 and 3 for the percentage of respondents who reported one or more instances of harm by gender/ academic position, and type of harm. See Table S1 in Supplementary Materials for the detailed results of the Mann-Whitney U tests.

Notably, there were few career stages in which there were no reports of harm. High proportions of assistant (50% of women and 26% of men) and associate professors (78% of women and 45% of men) reported experiencing personal harassment. Among gender diverse PhD students, 78% reported personal harassment. Women reported a disproportionate amount of harm – among post-docs, 13x the proportion of women compared to men reported personal harassment. Sexual harassment and assault are traumatic experiences and unacceptable in any situation, and against any person of any gender. It is particularly problematic that a higher proportion of women and gender diverse BSc and PhD students, and women MSc students and post-docs experienced sexual harassment and assault in their place of work or study compared to men. Combined with research showing that sexual harassment is negatively associated with retention (Leaper & Starr, 2019; Wilkins et al., 2023), these findings are deeply concerning to everyone interested in increasing diversity in physics.

**Figure 2**

*Percentage of Respondents Reporting One or More Instances of Experienced Harm by Gender/
Academic Position, and Type of Harm*

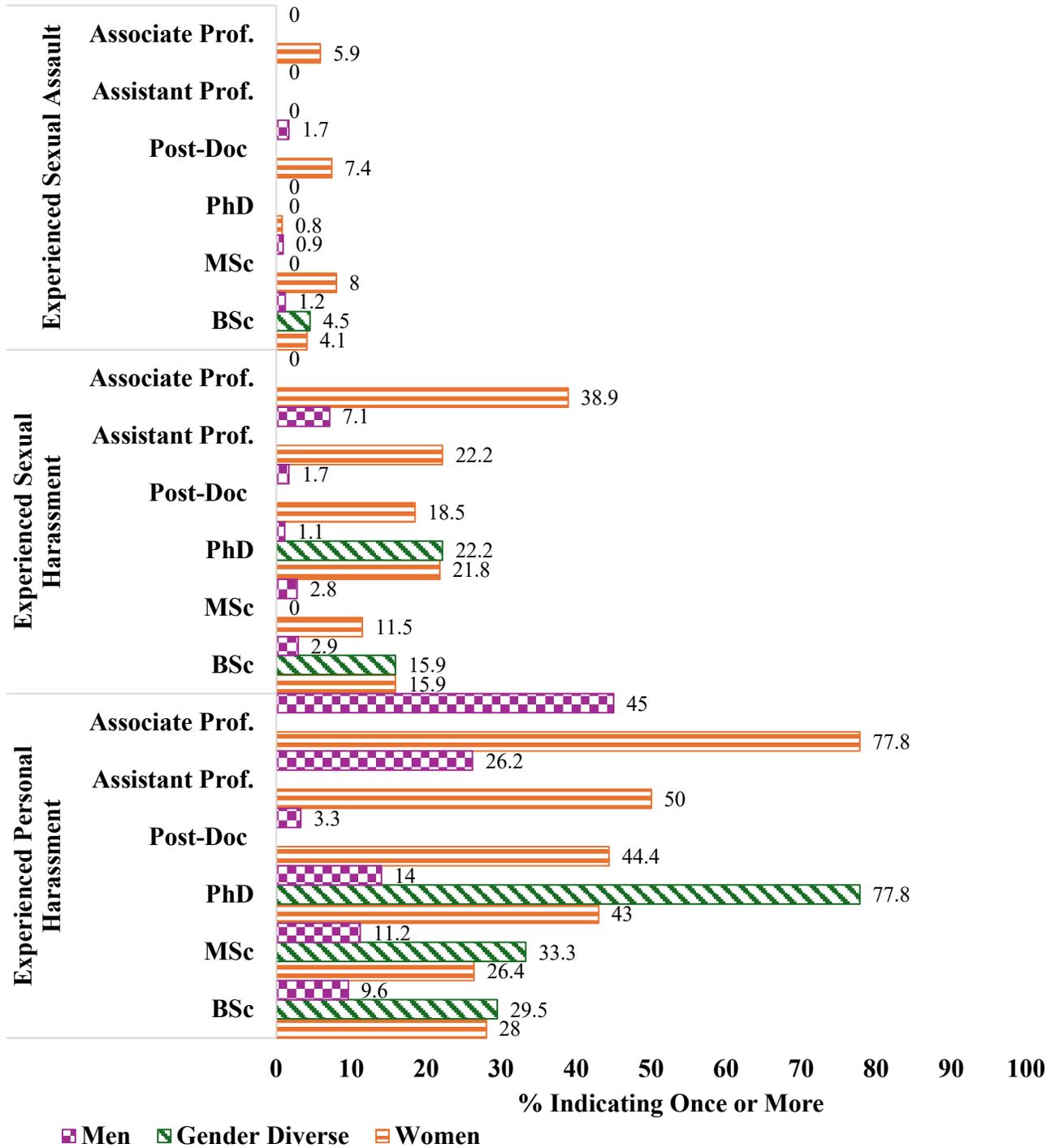

*Note*. Data for gender diverse assistant and associate professors is not reported due to sample sizes less than 5. There were no gender diverse post-docs in the sample. Missing bars indicate no respondents in that demographic group within position or numbers less than five. Missing bars and values of '0' indicate some presence of respondents, but all selected no instances of harm.

**Figure 3**

*Percentage of Respondents Reporting Awareness of One or More Instances of Harm by Gender/ Academic Position, and Type of Harm*

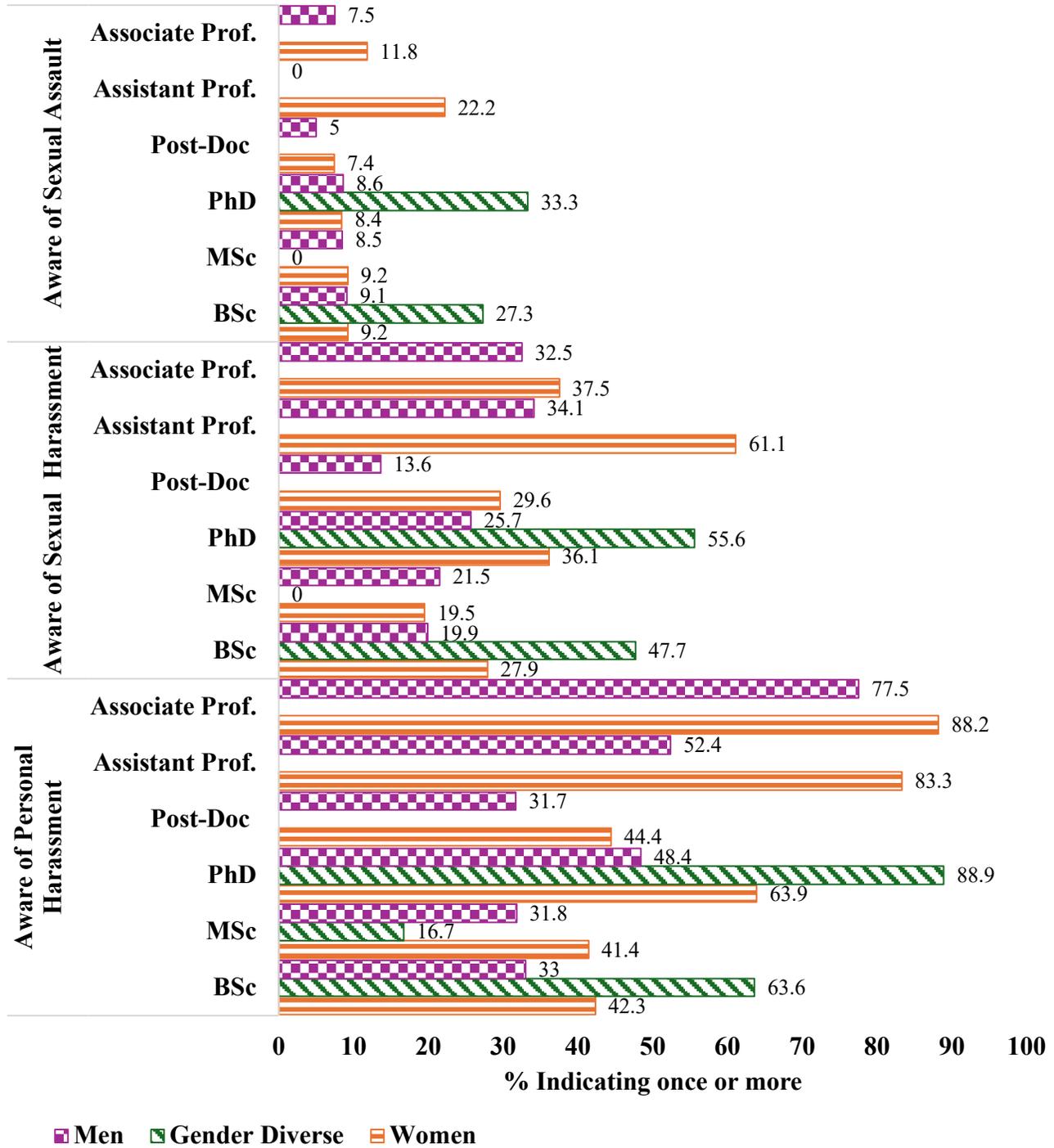

*Note*. Data for gender diverse assistant and associate professors is not reported due to sample sizes less than 5. There were no gender diverse post-docs in the sample. Missing bars indicate no respondents in that demographic group within position or numbers less than five. Missing bars and values of '0' indicate some presence of respondents, but all selected no instances of harm.

### Comparing Students and Professionals

Mann-Whitney U tests revealed that professionals reported more experiences of personal harassment ($p < .001$), awareness of personal harassment ($p < .001$), experiences of sexual harassment ($p = .05$), and awareness of sexual harassment ($p < .001$) than students. See Table S2 in Supplementary Materials.

## 3. How Did Frequencies of Experienced Harm and Awareness of Harm Differ Across Binary Gender Identities and Race?

Consistent with previous research (Aldosary et al., 2023; Clancy et al., 2017; Paradis et al., 2023; Wilkins et al., 2023), we found that gender was an important factor in understanding harmful experiences in physics education and workplaces. However, previous research has demonstrated the importance of disaggregating data by various types of identities (e.g., Clancy et al., 2017; Wilkins et al., 2023), so we next assessed the extent to which experiences of harm differed across the multiplicative dimensions of gender and race.

Contrary to expectations, analyses revealed that gender differences in frequencies of experienced harm were consistent across racial groups, such that BIPOC women and White women reported similar frequencies of experienced personal harassment, sexual harassment, and sexual assault ($ps > .05$), and both groups of women reported significantly higher frequencies of these experiences than BIPOC men and White men ($ps < .01$). BIPOC men and White men did not differ from each other on any of these experiences ($ps = 1.00$). Table 5 contains mean frequencies and standard deviations; see Supplementary Materials for detailed results of the Kruskal-Wallis tests. See Figure 4 for the proportions of each gender/racial group who reported one or more instances of each type of experienced harm.

Analyses revealed statistically significant group differences on awareness of personal harassment by race and gender; that is, the extent to which respondents were aware of others who had experienced harm. BIPOC women, White women, and White men all reported similar frequencies of awareness of personal harassment *(ps > .05)* that were significantly higher than BIPOC men *(ps < .01)*. Note however, that we used the Bonferroni correction, which is conservative. Using the unadjusted p-values, BIPOC women (*p* = .06) and White women (*p* = .01) reported higher frequencies than White men. In other words, women reported the highest awareness of personal harassment, followed by White men, and BIPOC men reported the lowest frequency. White women and BIPOC women reported similar frequencies of awareness of sexual harassment (*p* > .05; unadjusted *p* = .04, such that White women reported more), and White women reported statistically significantly more than BIPOC men and White men (*ps* < .01). Stated differently, White women reported the most frequent awareness of sexual harassment, and BIPOC men reported the least, with BIPOC women and White men in-between these groups. There were no group differences on awareness of sexual assault (*ps* >.05). See Figure 5 for the proportions of each gender/race group who reported awareness of each type of harm at least once or more. As shown in the figure, a substantial proportion of BIPOC gender diverse respondents, followed by White gender diverse respondents, reported awareness of harm, although these comparisons were not statistically tested.

**Table 5**

*Means and Standard Deviations for Types of Harm across Race/Gender*

| | Men | | | | Women | | | | Gender Diverse | | | | Total Sample | | | |
| | BIPOC ($n = 294$) | | White ($n = 839$) | | BIPOC ($n = 213$) | | White ($n = 508$) | | BIPOC ($n = 18$) | | White ($n = 53$) | | BIPOC ($n = 530$) | | White ($n = 1410$) | |
| | *M* | *SD* | *M* | *SD* | *M* | *SD* | *M* | *SD* | *M* | *SD* | *M* | *SD* | *M* | *SD* | *M* | *SD* |
|---|---|---|---|---|---|---|---|---|---|---|---|---|---|---|---|---|
| Experienced Personal Harassment | 1.26 | .61 | 1.28 | .68 | 1.73 | 1.00 | 1.66 | 1.02 | 1.50 | .79 | 1.75 | 1.07 | 1.46 | .83 | 1.43 | .85 |
| Awareness of Personal Harassment | 1.65 | .94 | 1.85 | .98 | 2.04 | 1.14 | 2.02 | 1.09 | 2.56 | 1.25 | 2.29 | 1.17 | 1.84 | 1.06 | 1.93 | 1.04 |
| Experienced Sexual Harassment | 1.04 | .22 | 1.05 | .27 | 1.26 | 1.67 | 1.37 | .81 | 1.28 | .67 | 1.25 | .63 | 1.14 | .49 | 1.17 | .57 |
| Awareness of Sexual Harassment | 1.31 | .68 | 1.44 | .77 | 1.46 | .84 | 1.62 | .96 | 2.22 | 1.31 | 1.86 | 1.11 | 1.41 | .79 | 1.52 | .87 |
| Experienced Sexual Assault | 1.00 | .06 | 1.01 | .13 | 1.09 | .47 | 1.07 | .38 | 1.11 | .47 | 1.02 | .14 | 1.04 | .32 | 1.03 | .25 |
| Awareness of Sexual Assault | 1.12 | .44 | 1.16 | .53 | 1.17 | .54 | 1.19 | .61 | 2.00 | 1.33 | 1.34 | .77 | 1.17 | .56 | 1.18 | .57 |

*Note.* Sample sizes (n's) may vary as some participants skipped certain questions.

**Figure 4**

*Percentage of Respondents Reporting One or More Instances of Experienced Harm by Race/Gender and Types of Harm*

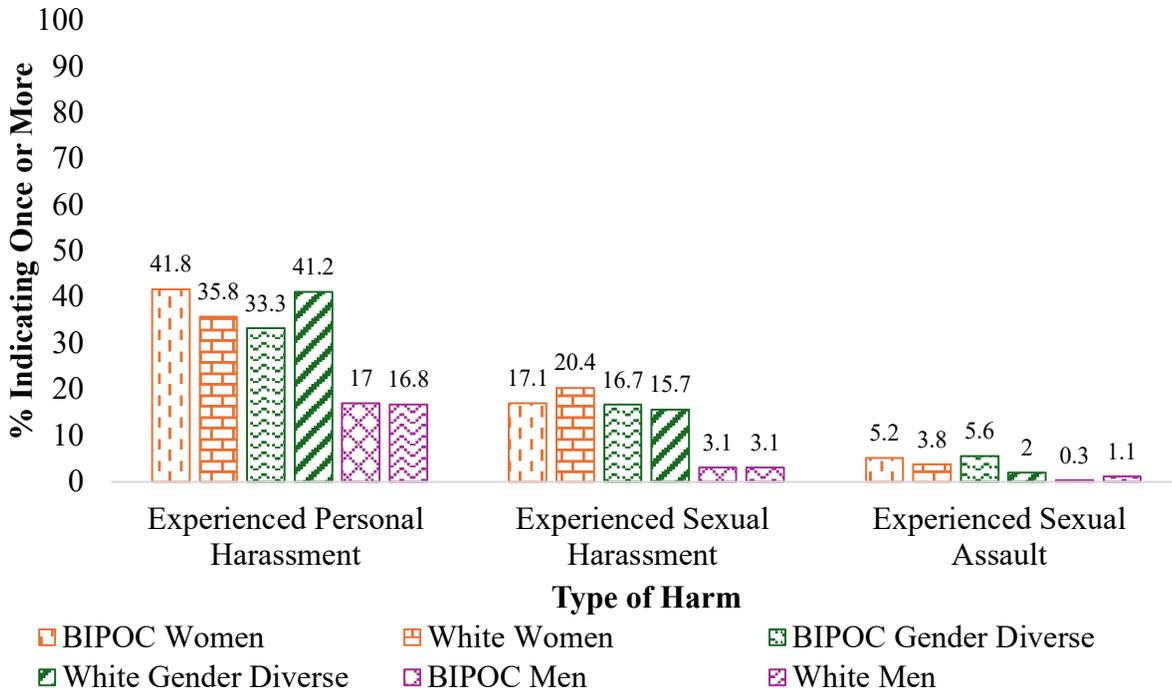

**Figure 5**

*Percentage of Respondents Reporting Awareness of One or More Instances of Harm by Race/Gender and Types of Harm*

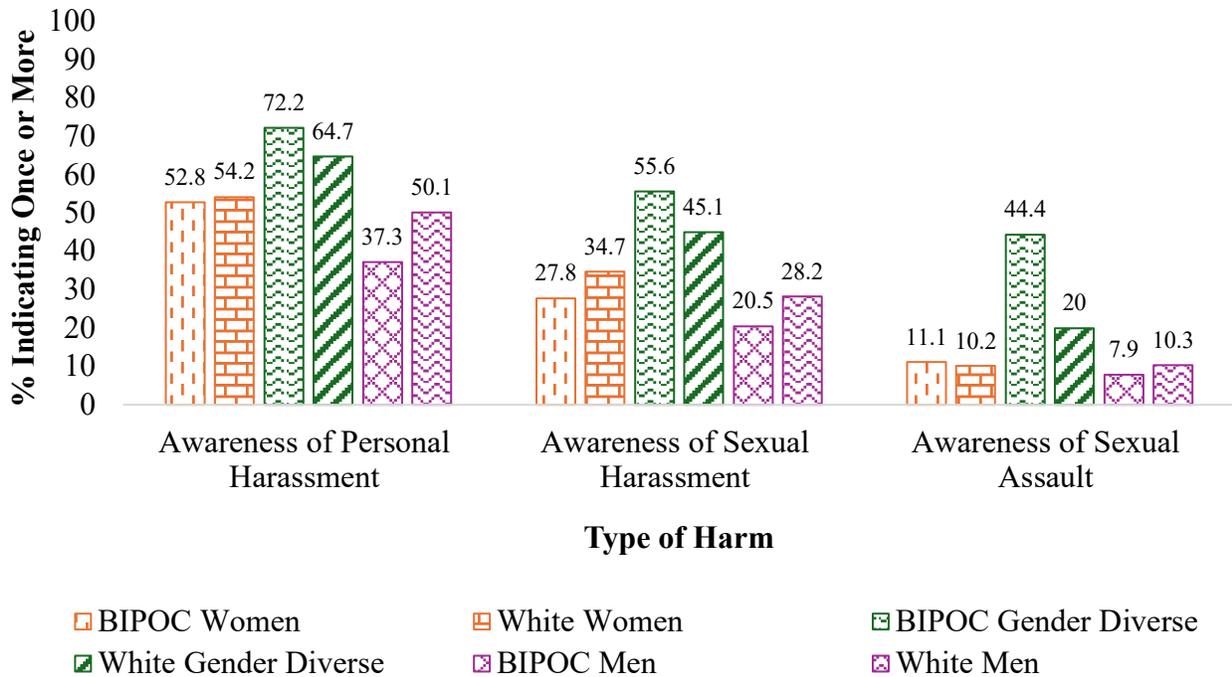

*Further Analysis of Black, Indigenous, and Persons of Colour Respondents*

We did not find statistical support for our expectations that race would interact with gender identity to predict more harmful experiences. However, this may be an artifact of the decision to combine Black, Indigenous, and Persons of Colour into one group. Although this was done to maintain statistical power with smaller samples, experiences may vary widely between these groups (Cech, 2022; Eom et al., 2025; Lee et al., 2022; Ovink et al., 2022; Shu et al., 2022; Wilkins-Yel et al., 2019). Therefore, to further explore the effect of race while maintaining maximum possible sample sizes, we combined Black and Indigenous respondents (including those who were multi-racial) into one group and compared them to POC (all racialized respondents who did not identity as Black or Indigenous) and White respondents.

Using KW tests with these three groups, we found no differences between racial groups among women (all ps >.05). This may be a result of low sample size; combining Black and Indigenous women who responded to these questions, we reached a total sample size of only 21. Among men, we found some evidence that Black and Indigenous respondents experienced more personal harassment than POC respondents (unadjusted $p = .03$) and White respondents (unadjusted $p = .05$). Note however, that these are p-values that have not been adjusted for multiple comparisons, and that the omnibus test did not reach statistical significance ($p = .10$), so these pairwise comparisons should be interpreted accordingly. Next, both Black and Indigenous men ($p = .05$) and White men ($p < .001$) reported more awareness of personal harassment than POC men. There was some evidence that Black and Indigenous men reported more sexual harassment than POC men (unadjusted $p = .02$) and White men (unadjusted $p = .03$). Note however, that the omnibus test did not reach significance ($p = .07$). Both Black and Indigenous men ($p = .05$) and White men ($p = .005$) reported more awareness of sexual harassment than POC men. There were no group differences on experienced sexual assault or awareness of sexual assault. See Supplementary Materials for detailed results of these tests.

To move our inquiry beyond null hypothesis significance testing alone, we disaggregated data by racial groups and found descriptive differences in experiences of Black, Indigenous, POC, and White respondents that may have been obscured by aggregation; see Table 6. Descriptively, Black women were most likely to experience personal harassment – about 1.5x the proportion of Black women compared to White women reported experiencing personal harassment at least once. The group with the highest proportion of experienced sexual harassment was Indigenous women, and the group with highest proportion of experienced sexual assault was Black women. A similar pattern of results was observed among men. About 2x the proportion of Black men compared to White men reported experiencing personal harassment at least once. Among men, the group with the highest proportion of those who experienced sexual harassment was Indigenous men; almost 5x the proportion of White men who experienced sexual harassment. However, the sample sizes of Black and Indigenous respondents were very small, so results should be interpreted on a descriptive basis. Simultaneously, this pattern calls on the physics community to prioritize more robust data collection, particularly among the least represented physicists. These findings indicate that race matters for experiences in physics, but that experiences may differ depending on racial identity, with Black and Indigenous physicists experiencing more harm. However, to say that some groups experience more harm is not to trivialize the harm experienced by others. For example, a quote below illustrates an experience of both sexism and racism.

### Comments Illustrating Gender and/or Race and Experienced Harm

The following comments are examples of harm experienced by racialized respondents.

'The comment I heard was "You are too beautiful to be Chinese." It appears to be a compliment but turns out a racial comment [....] It is easy for one to think of the opposite side "is a Chinese female supposed to be ugly?" (participant 2268, woman).

'Been called racial slurs multiple times' (participant 779, man).

**Table 6**

*Frequencies for Each Type of Harm Across Partially Disaggregated Race/Gender*

| | | Women | | | | | Men | | | | |
| | | **Black** (*n* = 12) | **Indigenous** (*n* = 9) | **POC** (*n* = 197) | **BIPOC** (*n* = 213) | **White** (*n* = 508) | **Black** (*n* = 17) | **Indigenous** (*n* = 13) | **POC** (*n* = 265) | **BIPOC** (*n* = 294) | **White** (*n* = 839) |
|---|---|---|---|---|---|---|---|---|---|---|---|
| Experienced Personal Harassment | Never | 50.0% | 55.6% | 58.9% | 58.2% | 64.2% | 64.7% | 76.9% | 84.5% | 83.0% | 83.2% |
| | Once or more | 50.0% | 44.4% | 41.1% | 41.8% | 35.8% | 35.3% | 23.1% | 15.5% | 17.0% | 16.8% |
| Awareness of Personal Harassment | Never | 25.0% | 44.4% | 49.0% | 47.2% | 45.8% | 47.1% | 38.5% | 65.0% | 62.7% | 49.9% |
| | Once or more | 75.0% | 55.6% | 51.0% | 52.8% | 54.2% | 52.9% | 61.5% | 35.0% | 37.3% | 50.1% |
| Experienced Sexual Harassment | Never | 83.3% | 66.7% | 83.1% | 82.9% | 79.6% | 94.1% | 84.6% | 97.7% | 96.9% | 96.9% |
| | Once or more | 8.3% | 33.3% | 16.9% | 17.1% | 20.4% | 5.9% | 15.4% | 2.3% | 3.1% | 3.1% |
| Awareness of Sexual Harassment | Never | 66.7% | 66.7% | 72.5% | 72.2% | 65.3% | 64.7% | 53.8% | 81.7% | 79.5% | 71.8% |
| | Once or more | 33.3% | 33.3% | 27.5% | 27.8% | 34.7% | 35.3% | 46.2% | 12.3% | 20.5% | 28.2% |
| Experienced Sexual Assault | Never | 91.7% | 100% | 94.4% | 94.8% | 96.2% | 100% | 100% | 99.6% | 99.7% | 98.9% |
| | Once or more | 8.3% | 0% | 5.6% | 5.2% | 3.8% | 0% | 0% | .4% | 0.3% | 1.1% |
| Awareness of Sexual Assault | Never | 91.7% | 100% | 88.0% | 88.9% | 89.8% | 88.2% | 84.6% | 92.7% | 92.1% | 89.7% |
| | Once or more | 8.3% | 0% | 12.0% | 11.1% | 10.2% | 11.8% | 15.4% | 7.3% | 7.9% | 10.3% |

**Notes.** Sample sizes (n's) may vary as some participants skipped certain questions.

Those who reported multiple racial identities (e.g., Indigenous and Person of Colour) were included in both categories in this table.

**4. How Did Frequencies of Experienced Harm and Awareness of Harm Differ Across Binary Gender Identities and Disability?**

Recall that we expected respondents with intersecting underrepresented identities to report greater experiences of harm. Supporting this expectation, analyses revealed that disabled women reported statistically greater frequencies of experienced personal harassment and sexual assault compared to all other groups, including women without disabilities ($ps < .05$). Women without disabilities reported significantly more frequent personal harassment and sexual assault than men without disabilities ($ps <.001$) but did not differ statistically from disabled men ($ps > .05$). Men with and without disabilities did not differ from each other on frequency of reported experiences of personal harassment or sexual assault ($ps > .05$). Note however, that we used the Bonferroni correction, which is conservative, and the smaller sample size of disabled men compared to men without disabilities may have reduced statistical power. Using the unadjusted p-values, disabled men reported a higher frequency of personal harassment experiences than men without disabilities ($p = .02$). In sum, our analysis of gender and disability on harmful experiences showed that disabled women reported the most personal harassment and sexual assault versus other groups, and men without disabilities reported the least personal harassment versus other groups. Women without disabilities and disabled men were in-between those groups. This pattern of findings suggests that gender and disability are relevant factors explaining the variability in experiences of personal harassment among those studying and working in physics in Canada.

Contrary to expectations, there were no differences between those with and without disabilities for experienced sexual harassment, however, women reported more frequent sexual harassment than men, regardless of disability ($ps < .01$). Again, we used the Bonferroni correction, which is conservative. Using unadjusted p-values, disabled women reported more sexual harassment than women without disabilities ($p = .05$). See Table 7 for mean frequencies and standard deviations, and Supplementary Materials for detailed results of the Kruskal Wallis tests. See Figure 6 for the proportions of gender/disability groups who reported one or more instances of each type of experienced harm. Notably, a high proportion of disabled women and gender diverse respondents reported harmful experiences.

Assessing awareness of harm, analyses revealed that disabled women, disabled men, and women without disabilities reported similar awareness of personal harassment ($ps > .05$). Women with and without disabilities reported significantly more awareness than men without disabilities ($ps < .01$). Although disabled men had the second highest mean rank score of the four groups, the difference between disabled men and men without disabilities was not statistically significant, likely due to small sample size that reduced statistical power (unadjusted p-value was $p = .02$). Overall, disabled women reported the most awareness of personal harassment versus other demographic groups, and men without disabilities reported the least. Disabled men and women without disabilities were in-between these groups.

For awareness of sexual harassment, women without disabilities reported significantly more harassment than men without disabilities ($p < .02$), underscoring the impact of gender on outcomes, but there were no other group differences ($ps > .05$). Although disabled women had the highest mean rank score, the difference between this group and men without disabilities was not statistically significant, likely due to small sample size that reduced the power of the test (unadjusted p-value, $p = .05$). Overall, disabled women reported the most frequent awareness of sexual harassment, followed by women without disabilities, disabled men, and men without disabilities. There were no group differences on awareness of sexual assault ($ps > .05$). Table 5 contains mean frequencies and standard deviations. See Supplementary Materials for detailed results of the Kruskal Wallis tests. Figure 7 contains proportions of each gender/disability group who reported being aware of one or more instances of each type of harm. As illustrated in Figure 7, a large percentage of gender diverse respondents and women with disabilities reported awareness of all three types of harm.

**Table 7**

*Means and Standard Deviations for Harm Variables by Disability/Gender*

| | Men | | | | Women | | | | Gender Diverse | | | | Total Sample | | | |
| | With Disabilities (*n* = 64) | | Without Disabilities (*n* = 973) | | With Disabilities (*n* = 61) | | Without Disabilities (*n* = 595) | | With Disabilities (*n* = 23) | | Without Disabilities (*n* = 32) | | With Disabilities (*n* = 150) | | Without Disabilities (*n* = 1616) | |
| | *M* | *SD* | *M* | *SD* | *M* | *SD* | *M* | *SD* | *M* | *SD* | *M* | *SD* | *M* | *SD* | *M* | *SD* |
|---|---|---|---|---|---|---|---|---|---|---|---|---|---|---|---|---|
| Experienced Personal Harassment | 1.52 | .94 | 1.24 | .63 | 2.10 | 1.32 | 1.64 | .96 | 2.38 | 1.20 | 1.25 | .57 | 1.87 | 1.19 | 1.39 | .79 |
| Awareness of Personal Harassment | 2.05 | 1.00 | 1.76 | .97 | 2.34 | 1.34 | 1.98 | 1.07 | 3.10 | 1.09 | 1.84 | .95 | 2.30 | 1.21 | 1.84 | 1.01 |
| Experienced Sexual Harassment | 1.05 | .21 | 1.04 | .26 | 1.53 | 1.05 | 1.31 | .74 | 1.76 | .94 | 1.03 | .18 | 1.35 | .82 | 1.14 | .51 |
| Awareness of Sexual Harassment | 1.39 | .75 | 1.38 | .74 | 1.66 | 1.03 | 1.53 | .90 | 2.57 | 1.33 | 1.41 | .76 | 1.66 | 1.03 | 1.44 | .80 |
| Experienced Sexual Assault | 1.02 | .13 | 1.01 | .12 | 1.20 | .65 | 1.06 | .36 | 1.10 | .44 | 1.00 | 0.0 | 1.10 | .46 | 1.03 | .24 |
| Awareness of Sexual Assault | 1.13 | .45 | 1.13 | .49 | 1.23 | .69 | 1.17 | .57 | 1.70 | 1.22 | 1.22 | .71 | 1.24 | .72 | 1.15 | .52 |

*Note.* Sample sizes (n's) may vary as some participants skipped certain questions.

**Figure 6**

*Percentage of Respondents Reporting One or More Instances of Experienced Harm by Disability/Gender and Type of Experienced Harm*

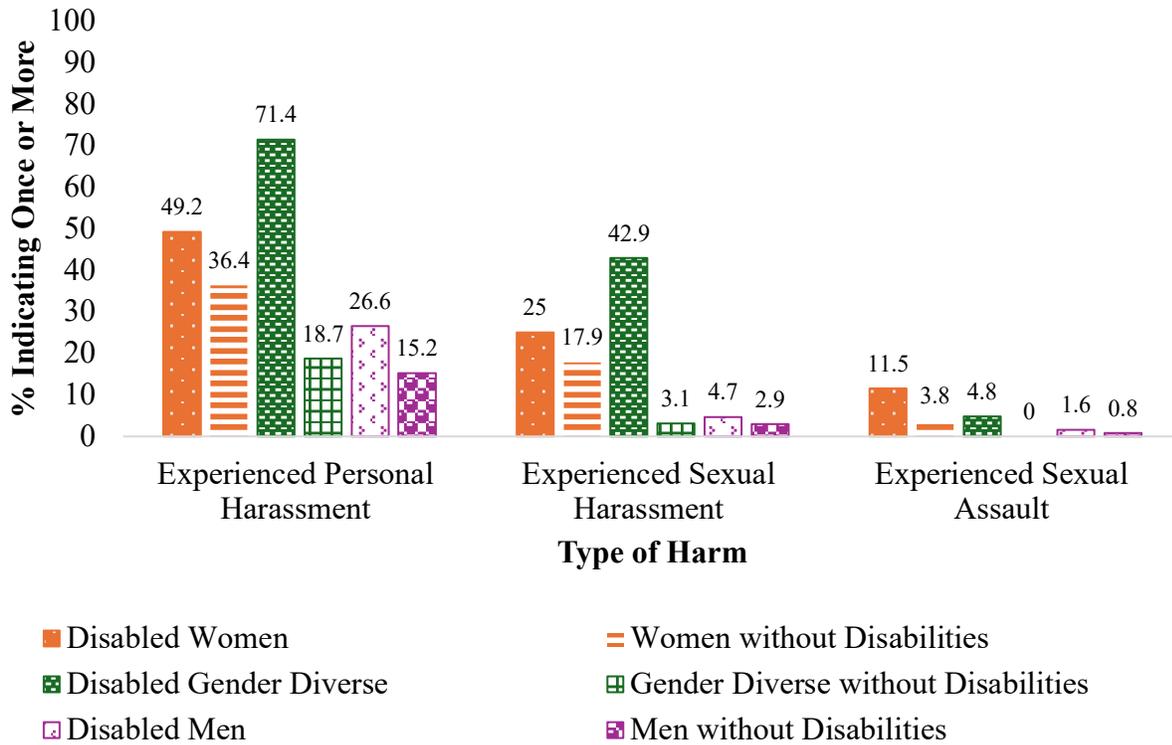

**Figure 7**

*Percentage of Respondents Reporting Awareness of One or More Instances of Harm by Disability/Gender and Type of Harm*

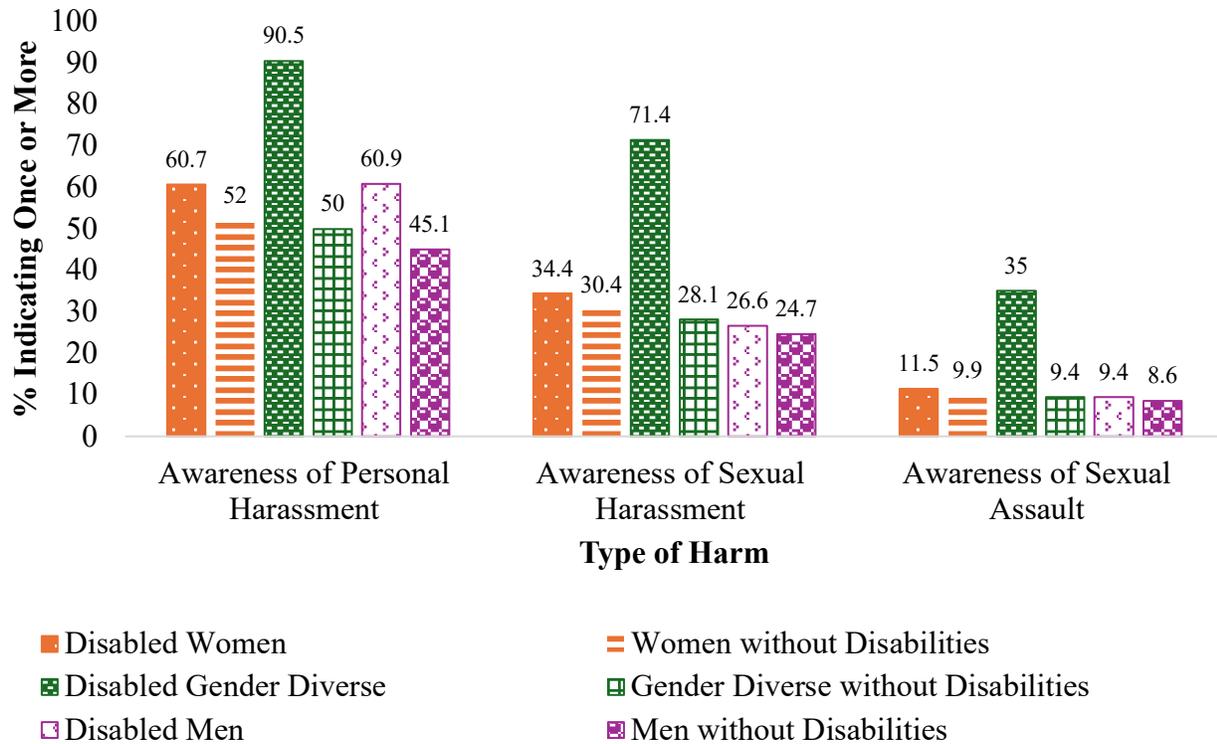

### Comments Illustrative of Disability and Experienced Harm

A comment shared by a participant with learning disabilities:

> 'Being told you're not fit to be a physicist by an undergrad advisor based on my learning disabilities and two poor course marks' (participant 201, man).

### Disability Accommodations

Disabled respondents (7.6% of the sample) were asked if their place of work or study provided accommodations. The majority (43.2%) reported that they did not seek out accommodations or did not need them, 27.8% reported receiving adequate accommodations, 17.6% reported receiving partial, inadequate accommodations, and 7.4% reported receiving full accommodations.

## 5. How Did Frequencies of Experienced Harm and Awareness of Harm Differ Across Binary Gender Identities and Sexual Orientation?

Women reported more frequent experiences of personal and sexual harassment than men, across heterosexual and sexually diverse respondents ($ps < .001$). Sexually diverse and heterosexual men did not differ from one another ($ps > .05$), and sexually diverse and heterosexual women did not differ from one another ($ps > .05$). However, sexually diverse women had a higher mean rank score for sexual harassment than heterosexual women. The small sample size of sexually diverse women may have reduced the statistical power of the test, and the unadjusted p-value was $p = .01$, suggesting that sexually diverse women experienced more sexual harassment than heterosexual women. Regarding sexual assault, sexually diverse and heterosexual women reported similar frequencies ($ps > .05$), and both reported more sexual assault than heterosexual men ($ps < .01$), but not more than sexually diverse men ($ps > .05$). Sexually diverse and heterosexual men did not differ from one another on sexual assault ($p = 1.00$). In other words, women reported the highest frequency of sexual assault, and heterosexual men reported the lowest frequency. Sexually diverse men reported levels in-between women and heterosexual men. See Table 8 for mean frequencies and standard deviations and Supplementary Materials for detailed results of the Kruskal Wallis tests. See Figure 8 for the proportions of each gender/sexual orientation group who reported one or more instances of each type of experienced harm. Notably, similar proportions of gender and sexually diverse respondents, sexually diverse women, and heterosexual women reported experienced harm.

Concerning awareness of harm, sexually diverse and heterosexual women reported similar awareness of personal harassment ($p = 1.00$), and both groups reported significantly more awareness than heterosexual men ($ps < .01$). Sexually diverse men reported similar frequencies to all other groups ($ps > .05$). Stated differently, women reported the most awareness of personal harassment, and heterosexual men reported the least. Sexually diverse men were in-between these groups. However, considering the mean rank scores, low statistical power due to small and unequal sample sizes could obscure differences between sexually diverse men and sexually diverse women (unadjusted p-value is $p$

= .04). Sexually diverse women reported the most awareness of sexual harassment; significantly more so than heterosexual men ($p$ = .01), followed by heterosexual women who also reported more awareness of sexual harassment than heterosexual men ($p$ = .06, unadjusted p-value $p$ = .009). There were no other group differences on awareness of sexual harassment. There were no group differences on awareness of sexual assault. See Table 8 for mean frequencies and standard deviations and Supplementary Materials for detailed results of the Kruskal Wallis tests. See Figure 9 for the proportions of each gender/sexual orientation group who reported awareness of one or more instances of each type of harm.

**Table 8**

*Means and Standard Deviations for Harm Variables by Sexual Orientation/Gender*

| | Men | | | | Women | | | | Gender Diverse | | | | Total Sample | | | |
|---|---|---|---|---|---|---|---|---|---|---|---|---|---|---|---|---|
| | Sexually Diverse (n = 122) | | Heterosexual (n = 1001) | | Sexually Diverse (n = 231) | | Heterosexual (n = 488) | | Sexually Diverse (n = 65) | | Heterosexual (n = 0) | | Sexually Diverse (n = 422) | | Heterosexual (n = 1489) | |
| | M | SD | M | SD | M | SD | M | SD | M | SD | M | SD | M | SD | M | SD |
| Experienced Personal Harassment | 1.22 | .57 | 1.28 | .68 | 1.68 | 1.04 | 1.68 | 1.00 | 1.70 | 1.03 | - | - | 1.55 | .95 | 1.41 | .82 |
| Awareness of Personal Harassment | 1.78 | .92 | 1.79 | .98 | 2.08 | 1.17 | 2.00 | 1.08 | 2.45 | 1.18 | - | - | 2.06 | 1.12 | 1.86 | 1.02 |
| Experienced Sexual Harassment | 1.02 | .13 | 1.05 | .27 | 1.38 | .80 | 1.32 | .77 | 1.28 | .65 | - | - | 1.27 | .67 | 1.14 | .51 |
| Awareness of Sexual Harassment | 1.42 | .74 | 1.40 | .75 | 1.63 | .98 | 1.55 | .91 | 2.03 | 1.18 | - | - | 1.63 | .97 | 1.45 | .81 |
| Experienced Sexual Assault | 1.02 | .13 | 1.01 | .12 | 1.08 | .43 | 1.08 | .41 | 1.05 | .28 | - | - | 1.06 | .35 | 1.03 | .25 |
| Awareness of Sexual Assault | 1.22 | .58 | 1.14 | .50 | 1.18 | .57 | 1.18 | .59 | 1.56 | 1.01 | - | - | 1.25 | .69 | 1.15 | .53 |

*Notes.* Sample sizes (n's) may vary as some participants skipped certain questions.

Although our sample did not contain any gender diverse respondents who identified as heterosexual, each person's sense of gender and sexuality are individually determined and limitless - some gender diverse people are heterosexual even though they are not represented in this sample.

**Figure 8**

*Percentage of Respondents Reporting One or More Instances of Experienced Harm by Sexual Orientation/Gender and Type of Harm*

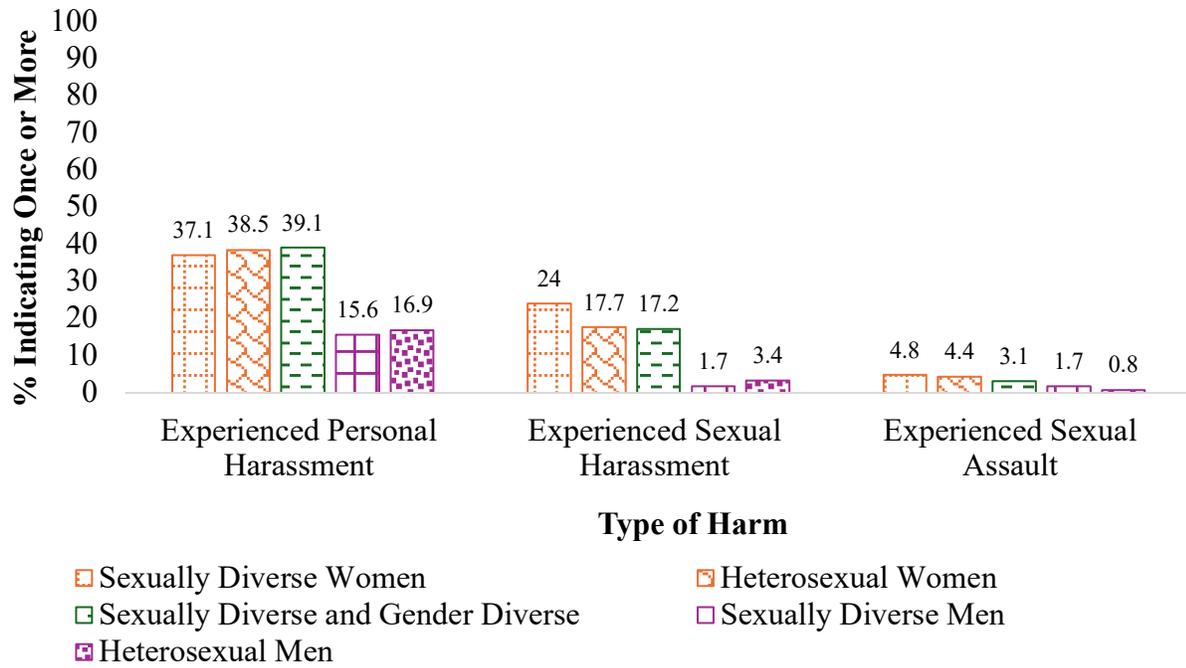

**Figure 9**

*Percentage of Respondents Reporting Awareness of One or More Instances of Harm by Sexual Orientation/Gender and Type of Harm*

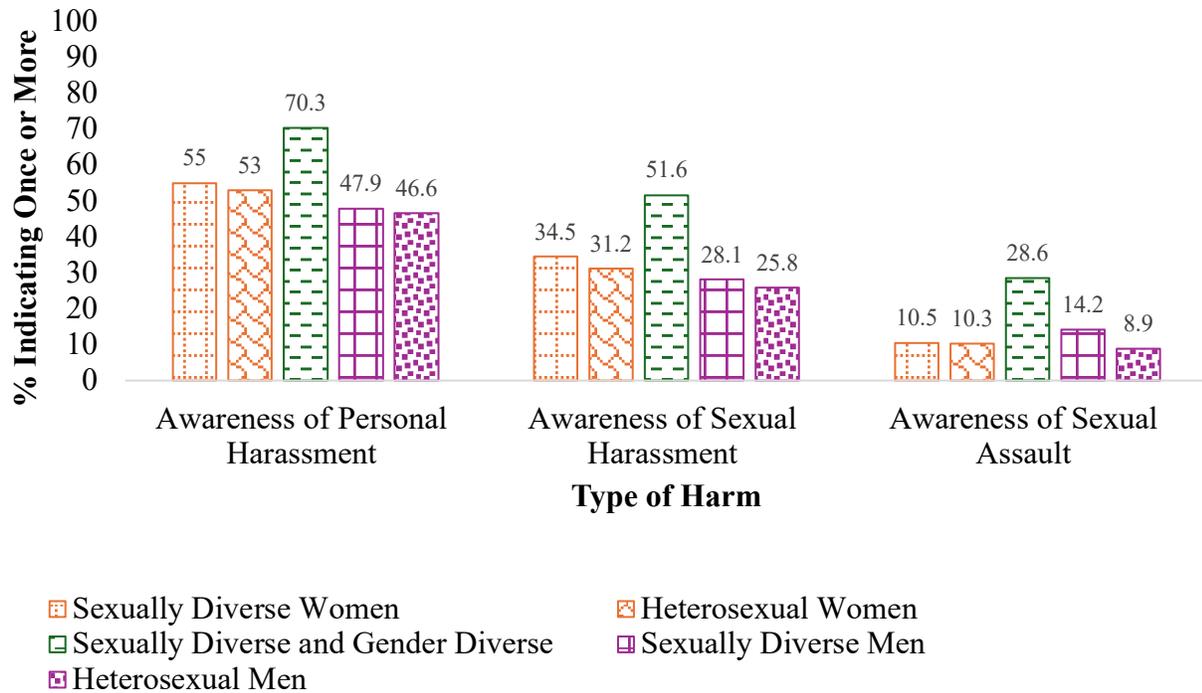

**Discussion**

    In a national survey, we asked those studying and working in physics about their experiences of harm – on a spectrum ranging from personal harassment to sexual assault. We applied an intersectional approach to analysis, assessing how experiences varied across gender identities, racial groups, disability, and sexual orientations. While much of the previous research on harmful experiences in physics has focused on U.S. contexts and singular identity dimensions (i.e., just binary gender; Aldosary et al., 2023; Aycock et al., 2019; Paradis et al., 2023), our findings provide a nuanced assessment of harmful experiences in the physics community in Canada. Although there are unlimited intersections of identity, we focused on gender identity and its intersections with race, disability, and sexual orientation due in part to practical limitations of sample size. We think it is imperative, however, for future research to gather larger sample sizes and utilize large-scale specific recruitment so that more understudied intersections (e.g., race and disability; race, sexual orientation, and gender) can be

considered with sufficient statistical power. Future research can also focus on qualitative methods to better understand experiences. However, the present analyses still can offer insights that are compatible with existing literature assessing these intersections of identities in physics (e.g., Barthelemy et al., 2022; Brown & Moloney, 2019; Clancy et al., 2017).

What our findings revealed may not be surprising, such that in general, awareness of harm or exposure to others' experiences of harassment or assault was more prevalent than directly experienced harm. Indeed, a single incident of harm can impact multiple witnesses or confidants. However, the fact that over half of respondents reported awareness of personal harassment at least once is concerning, especially given that witnessing harm can have significant psychological and health effects (Acquadro Maran et al., 2022; Nielsen et al., 2024). This finding underscores that institutional responses to harmful experiences should extend beyond those directly involved to consider the impact on those affected through being made aware of harm whether this is through witnessing harm, hearing a disclosure of harm, or being made aware some other way. Importantly, we found that majority-group members were aware of harm occurring in their places of work and study, supporting work by Dancy and Hodari (2023).

Although less common than awareness of harm happening to others, direct experiences of harm were also reported. Approximately one in four respondents reported experiencing personal harassment, and nearly one in ten reported experiencing sexual harassment in their place of work or study. Although these figures are lower than those reported in other studies (e.g., Aldosary et al., 2023; Aycock et al., 2019; Paradis et al., 2023; Starr & Leaper, 2019; Wilkins et al., 2023), any harmful experience is unacceptable and may influence attrition from physics. As prior research has demonstrated, harassment is associated with decreased retention and career progression in STEM (Clancy et al., 2017; Park et al., 2020; Leaper & Starr, 2019; Litzellachner et al., 2024; Wilkins et al., 2023). In the context of our data, it is possible that harmful experiences occurring in the Canadian physics eco-system contribute to attrition. In this research, there was not a single identity group who reported no harmful experiences, so working to reduce harm and improve the climate in physics will benefit everyone.

In other work based on this survey (Hennessey & Smolina et al., 2025), we found that women, BIPOC individuals (particularly Black and Indigenous people), and people with disabilities were underrepresented in the Canadian physics landscape. In the current work we found evidence that those with underrepresented identities in physics, particularly those with multiple underrepresented identities, experienced more harm. High rates of harm were also reported by students and early career researchers meaning that physics may be losing diverse perspectives before careers advance.

Gender differences emerged consistently for both experienced harm and awareness of harm. We found a pattern across outcomes, such that women and gender-diverse respondents reported higher frequencies of personal and sexual harassment compared to men, and women reported more sexual assault than men. These findings replicate patterns in STEM literature (Barthelemy et al., 2022; Clancy et al., 2017; Paradis et al., 2023; Wilkins et al., 2023). However, contrary to expectations, there were no statistically significant differences between women and gender-diverse respondents in frequency of reported experiences of harm. This contrasts with other research suggesting that gender-diverse individuals are at heightened risk compared to cisgender counterparts (Barthelemy et al., 2022; Jaffray, 2020; Prokopenko & Hango, 2022; Wilkins et al., 2023) and may reflect contextual or methodological differences, and the small sample size of gender diverse respondents. Gender diverse respondents also reported more awareness of harm than men and women, both in frequency of experiences and in awareness of harm. This suggests that gender is a highly relevant identity dimension when assessing harmful experiences among physicists and warrants further investigation. It may be that due to higher risk of harm in daily life (Jaffray, 2020; Prokopenko & Hango, 2022), gender dierse individuals are more attuned to harm.

We expected that respondent race would interact with gender identity, such that BIPOC respondents who were also women or gender diverse would report the most harm. Our analysis provided inconsistent support for this expectation. Initial comparisons between White and BIPOC respondents did not yield significant differences in frequencies of experienced harm on any outcome. However, disaggregated analyses on a descriptive level revealed that larger proportions of Black and Indigenous respondents reported experiences of personal and sexual harassment compared to White respondents. These findings emphasize the need for future research to prioritize data disaggregation to uncover patterns that may be obscured by applying broad racial categories. Further, findings highlight the need to create programs and policies that recognize the unique experiences and needs of intersectional groups. When intersectionality is not considered, exclusion can be magnified. For example, Lundy-Wagner (2013) called attention to experiences of Black men in STEM who face exclusion based on race and are further excluded from conversations on inclusion due to the focus on women.

Cultural stereotypes and representation in physics may help explain why aggregating BIPOC individuals obscured differences. Groups overrepresented in physics and stereotyped as intellectually competent (e.g., White and Asian individuals; Shu et al., 2022) may be perceived as more "prototypical" scientists, potentially shielding them from certain types of harm. Conversely, those who are both underrepresented and negatively stereotyped may be viewed as "outsiders," increasing their exposure to exclusion and harassment. Illustrating how being perceived as an outsider can increase exposure to harm, Danbold and Huo (2017) found that men who more strongly believed that men were prototypical scientists (and that this status was deserved) felt more psychological threat when exposed to initiatives designed to increase women's representation in STEM, and in turn reported higher likelihood of engaging in discriminatory behaviors towards women. However, this explanation is speculative, and it is important not to minimize the harm experienced by all racialized people in STEM.

For example, although people of some Asian backgrounds may be positively stereotyped in STEM (Shu et al., 2022), this can still have negative psychological effects (e.g., Gupta et al., 2011; McGee et al., 2023), robs people of their humanity and individuality, reduces them to stereotypes, denies them access to help when it is needed (Shah, 2019), and puts excessive pressure on people to live up to the stereotypes (Park et al., 2025). Further, anti-Asian racism is still a common experience with negative consequences for sense of safety and mental health (e.g., Huynh et al., 2023).

Another potential explanation for the lack of statistically significant group differences by race is how people cope with discrimination. Racism is a stressful experience associated with deleterious consequences for mental and physical health (e.g., Paradies et al., 2015; Williams et al., 2019) and can be coped with similarly to other stressors, including by avoidance (i.e., cognitive strategies like denial; Jacob et al., 2023; West et al., 2009).  Repeated experiences of racism can lead to 'de-sensitization' wherein one downplays the severity of an experience to cope with it (Lewis et al., 2013; Ortiz, 2019). It is therefore possible that BIPOC respondents in this research underreported experiences of harm because avoidant coping made these experiences less salient, or because people did not wish to think about those experiences by reporting them in the survey. Future research should further explore how those with different identities cope with harmful experiences in physics so that reporting is accurate, and people can access support needed to cope with and address harm. Future research can use behavioral checklists (asking about specific experiences versus directly asking if one has ever experienced harassment/assault) to help reduce the potential impact of denial (Williams, 2016).

Our findings also highlight the importance of disability. Disabled women reported the highest rates of personal and sexual harassment and sexual assault. Disabled men also reported a higher frequency of experiences and awareness of personal harassment than men without disabilities, further supporting the relevance of disability in understanding experiences of exclusion in physics, inclusive of gender identity. Although we did not compare this statistically due to small sample sizes, an alarmingly high proportion of gender diverse respondents with disabilities reported experiences of harm and awareness of harm. These findings are consistent with intersectional theories of compounded marginalization and existing research on ableism in STEM (Guthrie et al., 2025; Sukhai & Mohler, 2017) and other contexts (Brown & Maloney, 2019). We combined all disabled participants into a single category; however, we note that this approach, used due to sample size restrictions, erases the unique experience of people with different types of disabilities. Disabilities vary in chronicity, visibility, and the extent to which they may impact participation in physics (for example, someone using a mobility device faces barriers to participating in an inaccessible laboratory; Jeannis et al., 2020).

Statistical differences in experienced personal harassment by sexual orientation did not emerge in this study, diverging from earlier research (e.g., Barthelemy et al., 2022; Wilkins et al., 2023). However, we did find descriptive evidence that sexually diverse women experienced more sexual harassment than heterosexual women, suggesting that intersectional experiences are still relevant. Like disability, sexual orientation may be invisible, and as such, could influence the likelihood of experiencing harm and/or one's willingness to report it.

Finally, patterns in awareness of harm mirrored those seen in direct experiences: those most likely to experience harm—particularly women, Black, Indigenous, and gender-diverse respondents, and people with disabilities—were also the most likely to report awareness of harm. These results support the notion that people who are more likely to experience harm themselves may be more attuned to, or more frequently exposed to, others' disclosures of harm, possibly due to a sense of shared community (Acquadro Maran, 2022). Interestingly, White men reported relatively high levels of awareness of harm compared to BIPOC men, which could be due to their overrepresentation in leadership roles where formal complaints are more commonly encountered.

Our findings underscore an urgent need to elicit systemic change within the physics community in Canada and call for institutions to recognize the inequitable ways in which harm impacts different demographic communities. We assert that institutional policies aimed at reducing discrimination should therefore: 1) be intersectional in design, acknowledging how gender, race, disability, and sexual orientation interact to shape experience; 2) be developed with input from individuals with lived experiences of harm; 3) include explicit support for those witnessing experiences, receiving disclosures, or are otherwise made aware of harm, in addition to those directly impacted; 4) provide education and resources to bolster community capacity for responding to discrimination and disclosures, and 5) prioritize data collection and disaggregation in efforts to monitor progress to uncover hidden disparities. However, data alone is often not sufficient to compel action, so we call on the physics community in Canada to mobilize and engage with processes that effectively attenuate harm. This represents a challenge, because those with the most represented identities and the most power are often the same people who do not recognize when harm occurs, and consciously or unconsciously work to protect their interests by maintaining the status quo. For example, Dancy (2025) interviewed White, cis-gendered men teaching physics courses and revealed that many of these instructors did not recognize sexism and racism in physics, and therefore, may not recognize it when it occurs in their own classrooms.

This lack of recognition of harm in the classroom is consistent with other research by Dancy and colleagues finding that White men in physics were largely unaware of the importance of race or gender on experiences in physics (Dancy et al., 2020). The same study found that over half of White women did not perceive any race differences in physics experiences. Other research not specific to STEM has found that men were less likely than women to recognize, and therefore confront sexism (Drury & Kaiser, 2014). Even White men in physics who perceive themselves as progressive may be motivated (consciously or not) to uphold their privileged position by denying inequity, using discourse that supports racist and sexist norms, and positioning the causes of inequity as something over which they have little control, justifying inaction (Dancy & Hodari, 2023). Given that White people comprise the numerical majority in physics in Canada (Hennessey & Smolina et al., 2025), there are countless missed opportunities for them to confront harm and to model effective collective action if they do not first acknowledge that it occurs.

It is therefore crucial that *all* physicists, but particularly those afforded more privilege by their identities and positions of power and influence, take the time to thoughtfully educate themselves on issues of inequity, stereotypes, and bias, and commit to taking action against racism, sexism, ableism, homophobia, transphobia, and any other forms of discrimination and oppression in physics, even when it is uncomfortable. Considering the work by Dancy and Hodari (2023), we implore readers to not relegate this work to others, but to examine their sphere of influence and identify actions they can take.

**Limitations and Future Directions**

This study has limitations that must be considered in interpreting the findings. Firstly, like much social research, respondents self-selected to participate in the survey which can introduce bias. Next, the data contained unequal group sizes which limited our ability to make some statistical comparisons. Although non-parametric methods were used where appropriate, statistical power remained a challenge and may account for lack of anticipated statistically significant group differences. However, the challenge inherent to recruiting larger samples of more diverse respondents is that there are very few, creating a cycle of underrepresentation in the data, and thus, potential underreporting of harmful experiences among the least represented.

A methodological limitation is that we employed single-item measures of harm. While on the one hand, this reduces survey length and increases completion rates, single-item measures are less reliable than multi-item scales (Allen et al., 2022b). Furthermore, four of the six single-item harm questions were about sexual harassment and assault. Although sexual harassment and assault are experienced by people of all genders, they are more commonly experienced by women and gender minorities (Statistics Canada, 2024). The gendered nature of the questions may have primed respondents to consider gendered experiences, perhaps leading to underreporting of experiences of racism, homophobia, ableism, and other forms of identity-based harm. Future research could expand the scope of questions to explicitly include a variety of harmful experiences such as microaggressions, which are often the most frequent form of harm experienced by the least represented people (e.g., Barthelemey et al., 2016; Corbett et al., 2024).

## Conclusion

Limitations considered, this research offers critical insights into the patterns and group differences in experienced harm and awareness of harm within the Canadian physics community. It highlights the disproportionate burden of harm experienced by those with the least represented identities; particularly women, gender-diverse people, Black and Indigenous people, racialized people, disabled people, and sexually diverse people. Awareness of harm, an often-overlooked aspect of the harassment literature, emerged as both common and disproportionately experienced by underrepresented respondents. Addressing harm in the physics community (and academia as a whole) requires more than policy or performative commitment; true cultural transformation takes time, coordination, resources, accountability, transparency, empathy, and the sustained will to change. A vital first step is creating a data-informed understanding of physicists' experiences, which can provide a solid foundation from which to construct a more inclusive physics community.

## Author Contribution Statement

Hewitt and Smolina conceptualized the project and developed the methodology. Hewitt is the PI on the project and wrote and obtained Research Ethics board approval from Dalhousie University under REB 2020-5261. Ghose is the PI on the NSERC CWSE grant and supervised the research analysis and manuscript preparation. Smolina developed the questions and survey instrument in both languages and translated French entries. Smolina and E. Hennessey conducted data curation. Tassone conceptualized the research questions, designed and conducted formal analysis, created charts and tables, and conducted the literature search. E. Hennessey conceptualized research questions, created charts, and conducted and supervised analyses. Tassone and E. Hennessey wrote the initial draft of the paper, with input from Smolina, Hewitt, and Ghose. S. Hennessey conducted data analysis and visualization.

## Inclusion and Ethics Statement

As a research team, we worked together across provinces and institutions as a demographically diverse group (i.e., diversity in gender, race, sexuality, and disability as well as discipline), in collaboration with the Canadian Association of Physicists (CAP). We cite national literature relevant to the Canadian context through Statistics Canada reports as well as research with relevance to particular underrepresented communities. The study was approved by the Dalhousie Research Ethics Board, and data were anonymized to protect respondent confidentiality.

## Data Availability Statement

The dataset generated by the survey research during and/or analyzed during the current study are not available because doing so would violate our Dalhousie Research Ethics Board approval.

## Acknowledgements

Without the willingness of the physics community to respond to the survey and share their experiences, this work would not exist. We thank all respondents for their time and candor.

We thank the Canadian Association of Physicists for their support of this project and advertising the survey to the Canadian physics community.

Hewitt acknowledges that his work is done in Kjipuktuk (Halifax), which is located in Mi'kma'ki, the ancestral and unceded territory of the Mi'kmaq People. This territory is covered by the Treaties of Peace and Friendship which Mi'kmaq Wəlastəkwiyik (Maliseet), and Passamaquoddy Peoples first signed with the British Crown in 1726. The treaties did not deal with the surrender of lands and resources but in fact, recognized Mi'kmaq and Wəlastəkwiyik (Maliseet) title and established the rules for what was to be an ongoing relationship between nations on Mi'kma'ki.

Smolina acknowledges that her work is done at The University of Toronto, the Hospital for Sick Children, and the Mouse Imaging Centre, which are situated on the traditional lands of the Huron-Wendat, the Seneca, and the Mississaugas of the Credit First Nation. These territories are protected by the Dish with One Spoon wampum agreement, which all guests of this land have a responsibility to honour and uphold.

The WinS team acknowledges that their work is done on the shared traditional territory of the Neutral, Anishnaabe and Haudenosaunee peoples. This land is part of the Dish with One Spoon Treaty between the Haudenosaunee and Anishnaabe peoples and symbolizes the agreement to share, protect our resources and not to engage in conflict. This project was support by the Natural Sciences and Engineering Research Council of Canada.

**References**

Acquadro Maran, D., Varetto, A., & Civilotti, C. (2022). Sexual harassment in the workplace:

consequences and perceived self-efficacy in women and men witnesses and non-witnesses.

*Behavioral Sciences*, *12*(9), 326. https://doi.org/10.3390/bs12090326

Aldosary, G., Koo, M., Barta, R., Ozard, S., Menon, G., Thomas, C. G., ... & Surry, K. (2023). A  first

look at equity, diversity, and inclusion of Canadian medical physicists: results from the 2021

COMP EDI climate survey. *International Journal of Radiation Oncology, Biology, and Physics*,

*116*(2), 305-313. https://doi.org/10.1016/j.ijrobp.2023.01.044

Allen, D., Dancy, M., Stearns, E., Mickelson, R., & Bottia, M. (2022a). Racism, sexism and disconnection: Contrasting experiences of Black women in STEM before and after transfer from community college. *International Journal of STEM Education*, *9*(1), 20.

 https://doi.org/10.1186/s40594-022-00334-2

Allen, M.S., Iliescu, D., & Greiff, S. (2022b). Single Item Measures in Psychological Science.

*European Journal of Psychological Assessment, 38*(1), 1-5.

 https://doi.org/10.1027/1015-5759/a000699.

Amaka, N. S. (2024). Intersectionality in education: addressing the unique challenges faced by girls of

colour in STEM pathways. *International Research Journal of Modern Engineering Technology,*

*and Science*, *6*(11), 3460. https://www.doi.org/10.56726/IRJMETS64288

American Institute of Physics (2020). *The time is now: Systemic changes to increase African-*

*Americans with bachelor's degrees in physics and astronomy.* The AIP National Task Force to

Elevate African American Representation in Undergraduate Physics and Astronomy (TEAM-

UP). Retrieved from: https://www.aip.org/sites/default/files/aipcorp/files/teamup-full-

report.pdf.

Aycock, L. M., Hazari, Z., Brewe, E., Clancy, K. B., Hodapp, T., & Goertzen, R. M. (2019). Sexual harassment reported by undergraduate female physicists. *Physical Review Physics Education Research*, *15*(1), 010121.

https://doi.org/10.1103/PhysRevPhysEducRes.15.010121

Barthelemy, R. S. (2020). LGBT+ physicists qualitative experiences of exclusionary behavior and harassment. *European Journal of Physics*, *41*(6), 065703. 10.1088/1361-6404/abb56a

Barthelemy, R. S., McCormick, M., & Henderson, C. (2016). Gender discrimination in physics and astronomy: Graduate student experiences of sexism and gender microaggressions. *Physical Review of Physics Education Research*, *12*(2), 020119. DOI:

https://doi.org/10.1103/PhysRevPhysEducRes.12.020119

Barthelemy, R. S., Swirtz, M., Garmon, S., Simmons, E. H., Reeves, K., Falk, M. L., ... & Atherton, T. J. (2022). LGBT+ physicists: Harassment, persistence, and uneven support. *Physical Review Physics Education Research*, *18*(1), 010124.

https://doi.org/10.1103/PhysRevPhysEducRes.18.010124

Brown, R. L., & Moloney, M. E. (2019). Intersectionality, work, and well-being: The effects of gender and disability. *Gender & Society*, *33*(1), 94-122. https://doi.org/10.1177/0891243218800636

Campbell L. G., Mehtani S., Dozier M. E., Rinehart J. (2013). Gender-heterogeneous working groups produce higher quality science. *PLOS ONE*, 8(10), Article e79147. https://doi-org.libproxy.wlu.ca/10.1371/journal.pone.0079147

Canada Research Chairs. (2025). *Equity, Diversity, and Inclusion Action Plan.* Government of Canada. Accessed March 16[th] 2026, https://www.chairs-chaires.gc.ca/program-programme/equity-equite/action_plan-plan_action-eng.aspx

Canadian Association of University Teachers (CAUT). (2018). Underrepresented and underpaid:

diversity and equity among Canada's post-secondary education teachers. Retrieved online from:

https://www.caut.ca/sites/default/files/caut_equity_report_2018-04final.pdf

Cech, E. A. (2022). The intersectional privilege of white able-bodied heterosexual men in STEM.

*Science advances*, *8*(24), eabo1558. 10.1126/sciadv.abo1558

Cech, E. A., Waidzunas, T. (2019). STEM Inclusion Study Organization Report: APS. Ann Arbour,

MI: University of Michigan.

Centre for Research Innovation and Support. (2024). Socio-Demographic Data Guide for Program

Evaluation. University of Toronto. Retrieved from:

https://cris.utoronto.ca/guides/socio-demographic-data/

Charleston, L. J., Adserias, R. P., Lang, N. M., & Jackson, J. F. (2014). Intersectionality and STEM:

The role of race and gender in the academic pursuits of African American women in STEM.

*Journal of Progressive Policy & Practice*, *2*(3), 273-293.

Clancy, K. B., Lee, K. M., Rodgers, E. M., & Richey, C. (2017). Double jeopardy in astronomy and planetary science: Women of color face greater risks of gendered and racial harassment.

*Journal of Geophysical Research: Planets*, *122*(7), 1610-1623.

https://doi.org/10.1002/2017JE005256

Coleman, B. R., & Yantis, C. (2024). Feeling a little uneasy: A comparative discourse analysis of

White and BIPOC college students' reflective writing about systemic racism. *Journal of

Social Issues*, *80*(2), 473-495. https://doi.org/10.1111/josi.12612

Corbett, E., Barnett, J., Yeomans, L., & Blackwood, L. (2024). "That's just the way it is":bullying and harassment in STEM academia. *International Journal of STEM Education*, *11*(1), 27.

https://doi.org/10.1186/s40594-024-00486-3

Corneille, M., Lee, A., Allen, S., Cannady, J., & Guess, A. (2019). Barriers to the advancement of women of color faculty in STEM: The need for promoting equity using an intersectional framework. *Equality, Diversity and Inclusion: An International Journal*, *38*(3), 328-348. Doi: 10.1108/EDI-09-2017-0199

Crenshaw, K. (1991). Race, gender, and sexual harassment. *s. Cal. l. Rev.*, *65*, 1467.

Danbold, F., & Huo, Y. J. (2017). Men's defense of their prototypicality undermines the success of women in STEM initiatives. *Journal of Experimental Social Psychology*, *72*, 57-66. https://doi.org/10.1016/j.jesp.2016.12.014

Dancy, M. (2025). Majority group physics instructors struggle to notice, name, and disrupt racist and sexist student interactions. *Physical Review Physics Education Research*, *21*(1), 010142. DOI: https://doi.org/10.1103/PhysRevPhysEducRes.21.010142

Dancy, M., & Hodari, A. K. (2023). How well-intentioned white male physicists maintain ignorance of inequity and justify inaction. *International Journal of STEM Education*, *10*(1), 45. https://doi.org/10.1186/s40594-023-00433-8

Dancy, M., Rainey, K., Stearns, E., Mickelson, R., & Moller, S. (2020). Undergraduates' awareness of White and male privilege in STEM. *International Journal of STEM Education*, *7*(1), 52. https://doi.org/10.1186/s40594-020-00250-3

Drury, B. J., & Kaiser, C. R. (2014). Allies against sexism: The role of men in confronting sexism. *Journal of social issues*, *70*(4), 637-652. https://doi.org/10.1111/josi.12083

Eom, D., Molder, A. L., Tosteson, H. A., Howell, E. L., DeSalazar, M., Kirschner, E., ... & Scheufele, D. A. (2025). Race and gender biases persist in public perceptions of scientists' credibility. *Scientific Reports*, *15*(1), 11021. https://doi.org/10.1038/s41598-025-87321-z

Feser, M. S., & Bauer, A. B. (2025). From belonging to success: evaluating the influence of sense of belonging to physics on first-year university students' academic outcomes and the benefits of a

holistic support program. *European Journal of Physics*, *46*(3), 035702. 10.1088/1361-6404/adca13.

Gale, S., Mordukhovich, I., Newlan, S., & McNeely, E. (2019). The impact of workplace harassment on health in a working cohort. *Frontiers in Psychology*, *10*, 1181. https://doi.org/10.3389/fpsyg.2019.01181

Gupta, A., Szymanski, D. M., & Leong, F. T. (2011). The "model minority myth": Internalized racialism of positive stereotypes as correlates of psychological distress, and attitudes toward help-seeking. *Asian American Journal of Psychology*, *2*(2), 101. Doi: 10.1037/a0024183

Guthrie, M. W., Wu, X., Scanlon, E. M., Syerson, E., Butler, J., Mora, B., ... & McPadden, D. (2025). "The system wasn't designed for us": Experiences of five disabled physics students. *The Physics Teacher*, *63*(5), 386-387. https://doi.org/10.1119/5.0273259

Hennessey, E. J.*, Smolina, A.*, Hennessey, S., Tassone, A., Jay, A., Ghose, S., & Hewitt, K. (2025). Canadian Physics Counts: An exploration of the diverse identities of physics students and professionals in Canada. *Facets*. (*joint first authors)

Hofstra B., Kulkarni V. V., Galvez S. M.-N., He B., Jurafsky D., McFarland D. A. (2020). The diversity–innovation paradox in science. *Proceedings of the National Academy of Sciences, USA*, 117(17), 9284–9291. https://doi.org/10.1073/pnas.1915378117

Hussain, M., & Jones, J. M. (2021). Discrimination, diversity, and sense of belonging: Experiences of students of color. *Journal of Diversity in Higher Education*, *14*(1), 63. http://dx.doi.org/10.1037/dhe0000117

Huynh, J., Chien, J., Nguyen, A. T., Honda, D., Cho, E. E., Xiong, M., ... & Ngo, T. D. (2023). The mental health of Asian American adolescents and young adults amid the rise of anti-Asian racism. *Frontiers in Public Health*, *10*, 958517. https://doi.org/10.3389/fpubh.2022.958517

Jackson, Z. A., Harvey, I. S., & Sherman, L. D. (2023). The impact of discrimination beyond sense of belonging: Predicting college students' confidence in their ability to graduate. *Journal of College Student Retention: Research, Theory & Practice24*(4), 973-987. DOI:10.1177/1521025120957601

Jacob, G., Faber, S. C., Faber, N., Bartlett, A., Ouimet, A. J., & Williams, M. T. (2023). A systematic review of Black people coping with racism: Approaches, analysis, and empowerment. *Perspectives on Psychological Science*, *18*(2), 392-415. https://doi.org/10.1177/17456916221100509

Jaffray, B.(2020). Experiences of violent victimization an unwanted sexual behaviors among gay, lesbian, bisexual, and other sexual minority people, and the transgender population, in Canada, 2018. Retrieved from: https://www150.statcan.gc.ca/n1/pub/85-002-x/2020001/article/00009-eng.htm

James, W., Bustamante, C., Lamons, K., Scanlon, E., & Chini, J. J. (2020). Disabling barriers experienced by students with disabilities in postsecondary introductory physics. *Physical Review Physics Education Research*, *16*(2), 020111. https://doi.org/10.1103/PhysRevPhysEducRes.16.020111

Jaramillo, A. M., Macedo, M., Oliveira, M., Karimi, F., & Menezes, R. (2025). Systematic comparison of gender inequality in scientific rankings across disciplines. *arXiv preprint arXiv:2501.13061*.

Jeannis, H., Goldberg, M., Seelman, K., Schmeler, M., & Cooper, R. A. (2020). Barriers and facilitators to students with physical disabilities' participation in academic laboratory spaces. *Disability and Rehabilitation: Assistive Technology*.https://doi.org/10.1080/17483107.2018.1559889

Jin, L., Zamudio, G., Wang, C.D., & Lin, S. (2024). Insecure attachment and eating disorder symptoms: Intolerance of uncertainty and emotion regulation as mediators. *Journal of Clinical Psychology, 80*(7), 1673-1688. Doi: 1002/jclp.23685.

Johnson, C. (2023). The purpose of whisper networks: a new lens for studying informal communication

   channels in organizations. Frontiers in Communication, 8.

   https://doi.org/10.3389/fcomm.2023.1089335

Jung, S., & Mendoza, J. (2023). A Gender-Based and Sexualized Violence Community Risk

   Assessment Tool for Post-Secondary Settings. Courage to Act: Addressing and Preventing

   Gender-Based Violence at Post-Secondary Institutions in Canada. Possibility Seeds.

   https://www.couragetoact.ca/cra.

Leaper, C., & Starr, C. R. (2019). Helping and hindering undergraduate women's STEM motivation:

   Experiences with STEM encouragement, STEM-related gender bias, and sexual harassment.

   *Psychology of Women Quarterly*, *43*(2), 165-183. https://doi.org/10.1177/0361684318806302

Lee, M. J., Collins, J. D., Harwood, S. A., Mendenhall, R., & Huntt, M. B. (2020). "If you aren't

   White, Asian or Indian, you aren't an engineer": racial microaggressions in STEM education.

   *International Journal of STEM Education*, *7*, 1-16.https://doi.org/10.1186/s40594-020-00241-4

Levandowski, B.A., Pro, G. C., Rietberg-Miller, S.B., & Camplain, R. (2024). We are complex beings:

   comparison of statistical methods to capture and account for intersectionality. *BJM open, 14(*1),

   e077194.

Lewis, J.A., Mendenhall, R., Harwood, S.A., *&* Browne-Huntt, M. (2013). Coping with Gendered

   Racial Microaggressions among Black Women College Students. *Journal of African American

   Studies, (17)*, 51–73. https://doi.org/10.1007/s12111-012-9219-0.

Litzellachner, L. F., Barnett, J., Yeomans, L., & Blackwood, L. (2024). How harassment is depriving

   universities of talent: a national survey of STEM academics in the UK. *Frontiers in Psychology*,

   *14*, 1212545. https://doi.org/10.3389/fpsyg.2023.1212545

Lundy-Wagner, V.,C. (2013). Is it Really a Man's World? Black Men in Science, Technology,

   Engineering, and Mathematics at Historically Black Colleges and Universities. *The *Journal of Negro Education, 82*(2), 157-168.

https://doi.org/10.7709/jnegroeducation.82.2.0157

Magoon, K., Robinson, M., Kissling, A., & Ozeua, V. (2022). Best Practices for Demographic Data

Collection & Reporting: Evaluator's Guide. Public Consulting Group. Retrieved from:

https://publicconsultinggroup.com/insight/best-practice-for-demographic-data-collection-

reporting-evaluators-guide/

Malcom, L., & Malcom, S. (2011). The double bind: The next generation. *Harvard Educational*

*Review, 81*(2), 162-172. https://doi.org/10.17763/haer.81.2.a84201x508406327

Mattheis, A., Marín-Spiotta, E., Nandihalli, S., Schneider, B., & Barnes, R. T. (2022). " Maybe this is

just not the place for me:" Gender harassment and discrimination in the geosciences. *PLoS*

*One*, *17*(5), e0268562.https://doi.org/10.1371/journal.pone.0268562

McGee, E. O., Thakore, B. K., & LaBlance, S. S. (2017). The burden of being "model": Racialized

experiences of Asian STEM college students. *Journal of Diversity in Higher Education, 10*(3),

253–270. https://doi.org/10.1037/dhe0000022

Meyer, J., Barnett, J., Corbett, E., Yeomans, L., & Blackwood, L. (2025). Increasing diversity in STEM

academia: a scoping review of intervention evaluations. *Studies in Higher Education*, 1-38.

https://doi.org/10.1080/03075079.2024.2442052

Minnotte, K. L., & Pedersen, D. E. (2023). Sexual harassment, sexual harassment climate, and the

well-being of STEM faculty members. *Innovative Higher Education*, *48*(4), 601

-618.https://doi.org/10.1007/s10755-023-09645-w

National Center for Science and Engineering Statistics (NCSES). 2023. Diversity and STEM: Women,

Minorities, and Persons with Disabilities 2023. Special Report NSF 23-315. Alexandria, VA:

National Science Foundation. Available at https://ncses.nsf.gov/wmpd.

Natural Sciences and Engineering Research Council of Canada. (2025a). *Dimensions: Equity, Diversity and Inclusion*. Government of Canada. Accessed March 16th 2026, https://nserc-crsng.canada.ca/en/equity-diversity-and-inclusion-9

Natural Sciences and Engineering Research Council of Canada. (2025b). *Current Chairs for Inclusion in Science and Engineering*. Government of Canada. Accessed March 16th 2026. https://nserc-crsng.canada.ca/en/current-chairs-inclusion-science-and-engineering

Nielsen, M. B., Einarsen, S. V., Parveen, S., & Rosander, M. (2024). Witnessing workplace bullying—A systematic review and meta-analysis of individual health and well-being outcomes. *Aggression and violent behavior*, *75*, 101908.https://doi.org/10.1016/j.avb.2023.101908

Ong, M., Wright, C., Espinosa, L., & Orfield, G. (2011). Inside the double bind: A synthesis of empirical research on undergraduate and graduate women of color in science, technology, engineering, and mathematics. *Harvard Educational Review, 81*(2), 172–208. http://doi.org/10.17763/haer.81.2.t022245n74752v2

Ortiz, S. M. (2019). "You Can Say I Got Desensitized to It": How Men of Color Cope with Everyday Racism in Online Gaming. *Sociological Perspectives*, *62*(4), 572–588. https://www.jstor.org/stable/26746202

Ovink, S. M., Byrd, W. C., Nanney, M., & Wilson, A. (2024). "Figuring out your place at school like this:" Intersectionality and sense of belonging among STEM and non-STEM college students. *PLoS One*, *19*(1), e0296389. https://doi.org/10.1371/journal.pone.0296389

Palid, O., Cashdollar, S., Deangelo, S., Chu, C., & Bates, M. (2023). Inclusion in practice: A systematic review of diversity-focused STEM programming in the United States. *International Journal of STEM Education*, *10*(2). https://doi.org/10.1186/s40594-022-00387-3

Paradies, Y., Ben J., Denson, N., Elias, A., Priest, N., Pieterse, A., Pieterse, A., Gupta, A.,

  Kelaher, M., & Gee, G.  (2015). Racism as a Determinant of Health: A Systematic

  Review and Meta-Analysis. *PLoS ONE 10*(9): e0138511. doi:10.1371/journal.pone.0138511

Paradis, K. C., Moran, J. M., & Hendrickson, K. R. (2023). Women in Medical Physics. Medical

  Physics, (Suppl, 1), 80-84. DOI: 10.1002/mp.16075

Park, A. S., Bahia, J., & Bing, A. (2025). "I was constantly being questioned": Racialized STEM

  Graduate Students in Canada. *Canadian Journal of Higher Education*, *55*(1), 55    -

  68.https://doi.org/10.47678/cjhe.v1i1.190355

Park, J. J., Kim, Y. K., Salazar, C., & Hayes, S. (2020). Student–faculty interaction and discrimination

  from faculty in STEM: The link with  retention. *Research in Higher Education*, *61*(3), 330-356.

  https://doi.org/10.1007/s11162-019-09564-w

Perreault, A., Franz-Odendaal, T., Langelier, E., Farenhorst, A., Mavriplis, C., & Shannon, L. (2018).

  *Analysis of distribution of gender in STEM fields in Canada*. Retrieved online from:

  https://ccwestt-ccfsimt.org/analysis-of-the-distribution-of-gender-in-stem-fields-in-canada/

Porter, A. M., & Ivie, R. (2019). Women in Physics and Astronomy, 2019. Report. *AIP Statistical

  Research Center*. Retrieved from: http://eric.ed.gov/?id=ED594227

Prokopenko, E., & Hango, D. (2022). Bullying victimization among sexually and gender diverse  youth

  in Canada. https://www150.statcan.gc.ca/n1/pub/75-006-x/2022001/article/00010-eng.htm

Purdie-Vaughns, V., & Eibach, R. P. (2008). Intersectional invisibility: The distinctive advantages and

  disadvantages of multiple subordinate-group identities. *Sex Roles*, *59*(5), 377-391.

  https://doi.org/10.1007/s11199-008-9424-4

Raj, A., Johns, N. E., & Jose, R. (2020). Gender parity at work and its association with workplace

  sexual harassment. *Workplace Health & Safety*, *68*(6), 279-292.

  https://doi.org/10.1177/2165079919900793

Rosa, K., & Mensah, F. M. (2016). Educational pathways of Black women physicists: Stories of experiencing and overcoming obstacles in life. *Physical Review Physics Education Research*, *12*(2), 020113. DOI: https://doi.org/10.1103/PhysRevPhysEducRes.12.020113

Rospenda, K. M., Richman, J. A., McGinley, M., Moilanen, K. L., Lin, T., Johnson, T. P., ... & Hopkins, T. (2023). Effects of chronic workplace harassment on mental health and alcohol misuse: a long-term follow-up. *BMC Public Health*, *23*(1), 1430. https://doi.org/10.1186/s12889-023-16219-0

Scarborough Charter. (2026). *Our Members.* Accessed March 16th 2026. https://scarboroughcharter.ca/membership/our-members/

Seyranian, V., Madva, A., Duong, N., Abramzon, N., Tibbetts, Y., & Harackiewicz, J. M. (2018). The longitudinal effects of STEM identity and gender on flourishing and achievement in college physics. *International journal of STEM education*, *5*(1), 40. https://doi.org/10.1186/s40594-018-0137-0

Shah, N. (2019). "Asians are good at math" is not a compliment: STEM success as a threat to personhood. *Harvard Educational Review*, *89*(4), 661-686. https://doi.org/10.17763/1943-5045-89.4.661

Shu, Y., Hu, Q., Xu, F., & Bian, L. (2022). Gender stereotypes are racialized: A cross-cultural investigation of gender stereotypes about intellectual talents. *Developmental Psychology*, *58*(7), 1345. https://doi.org/10.1037/dev0001356

Spencer, B. M. (2021, September). The psychological costs of experiencing racial discrimination in the ivory tower: The untold stories of Black men enrolled in science, technology, engineering, and mathematics (STEM) doctoral programs. In *Sociological Forum* (Vol. 36, No. 3, pp. 776-798).

Statistics Canada (2024). Gender Results Framework: A new data table on workplace harassment.

   Catalogue no. 11-001-X. Retrieved from: https://www150.statcan.gc.ca/n1/en/daily
   -quotidien/240212/dq240212a-eng.pdf?st=gvU3z1ND

Strickland, D. (2017). *Summary of CAP physics department survey – 2016.* Retrieved from:

   https://pic-pac.cap.ca/index.php/Issues/showpdf/article/v73n4.0-a3999.pdf

Sukhai, M. A., & Mohler, C. E. (2016). *Creating a Culture of Accessibility in the Sciences*. Academic

   Press.

Sulik, J., Bahrami, B., & Deroy, O. (2022). The diversity gap: when diversity matters for knowledge.

   *Perspectives on Psychological Science*, *17*(3), 752-767.

   https://doi.org/10.1177/17456916211006070

United Nations. *The right to access to and participate in science.* Accessed March 16[th] 2026.

   https://www.ohchr.org/en/special-procedures/sr-cultural-rights/right-access-and-participate-
   science#:~:text=The%20right%20to%20access%20to%2C%20to%20participate%20in%20and
   %20to,as%20engineering%2C%20technologies%20and%20health.

van Knippenberg, D. (2024). Team diversity and team performance: Paths to synergetic and disruptive

   effects. *Current Opinion in Psychology*, *59*, 101877.

   https://doi.org/10.1016/j.copsyc.2024.101877

West, L. M., Donovan, R. A., & Roemer, L. (2009). Coping With Racism: What Works and Doesn't

   Work for Black Women? *Journal of Black Psychology*, *36*(3), 331-349. https://doi-
   org.libproxy.wlu.ca/10.1177/0095798409353755

Williams, D. R. (2016). Improving the measurement of self-reported racial discrimination: Challenges

   and opportunities. https://doi.org/10.1037/14852-004

Williams, D. R., Lawrence, J. A., & Davis, B. A. (2019). Racism and health: evidence and needed

    research. *Annual review of public health*, *40*(1), 105-125. https://doi.org/10.1146/annurev-

    publhealth-040218-043750

Wilkins, K., Carroll, S. L., Davis, K. P., Hauptfeld, R., Jones, M. S., Larson, C. L., ... & Pejchar, L.

    (2023). Sexual harassment disproportionately affects ecology and evolution graduate students

    with multiple marginalized identities in the United States. *BioScience*, *73*(5), 376-387.

    https://doi.org/10.1093/biosci/biad032

Wilkins-Yel, K. G., Hyman, J., & Zounlome, N. O. (2019). Linking intersectional invisibility and

    hypervisibility to experiences of microaggressions among graduate women of color in STEM.

    *Journal of Vocational Behavior*, *113*, 51-60. https://doi.org/10.1016/j.vb.2018.10.018

Yang, Y., Tian, T. Y., Woodruff, T. K., Jones, B. F., & Uzzi, B. (2022). Gender-diverse teams produce

    more novel and higher-impact scientific ideas. https://doi.org/10.1073/pnas.2200841119

**Supplementary Material**

________________________________________________________________________

________________________________

<div align="center">

**Additional Demographic Information**

</div>

**Religion and Spirituality**

There was a diversity of religious and spiritual affiliations; 38% of participants were atheist, 16.8% reported no religious or spiritual affiliation, 15.5% were Christian, 15.1% were agnostic, 4.7% were Muslim, 1.6% were Hindu, 1.6% were Jewish, 1.2% were Buddhist, .4% were Sikh, .3% were an Indigenous spiritually, 4.2% were unsure or questioning, and .7% did not answer.

**Highest Education and Position**

Respondents came from a variety of educational backgrounds, since the survey was open to anyone perusing a physics degree or who already had a physics degree in Canada. 22.0% of respondents reported that their highest attained education was high school, 10.3% some university or college, 1.2% had a 1-2 year diploma, 20.3% had a 3-5 year bachelor's degree, 17.7% had a master's degree, 16.4% a PhD, 11.7% a post-doc, .3% had a professional degree (e.g., MD), .1% said other, and .1% preferred not to answer or did not answer.

Most respondents were students; 34.2% were undergraduates, and 26.5% were graduate students. 15.1% were working as faculty members, 4.5% were post-docs, 4.9% were working in research institutes, 8.1% in industry and government, 6.5% reported a different position, and .1% did not respond to this question.

<div align="center">

**Results of Statistical Tests**

</div>

**1. How did frequencies of experienced harm and awareness of harm differ across gender identities?**

To test differences in experienced and awareness of harm by gender (women, men, gender diverse), we used KW tests and Dunn post-hoc tests with the Bonferroni correction for multiple comparisons, as recommended by Ordak (2023). Effect sizes are eta-squared derived from the H statistic, as demonstrated by Tomczak and Tomczak (2014).

### Experienced Personal Harassment

A Kruskal-Wallis test indicated that there was a significant difference in experienced personal harassment across the three groups, H(2) = 114.99, $p$ <.001, $\eta^2$ = .06. The mean rank personal harassment score for gender diverse respondents was 1107.44, for women was 1090.95, and for men was 880.69. Dunn post-hoc tests with Bonferroni correction indicated that the mean rank personal harassment score for women was significantly higher than for men ($p$ < .001). The mean rank personal harassment score for gender diverse respondents was significantly higher than for men ($p$ < .001). However, there was no significant difference between women and gender diverse respondents ($p$ = 1.00).

### Experienced Sexual Harassment

A Kruskal-Wallis test indicated that there was a significant difference in experienced sexual harassment across the three groups, H(2) = 138.08, $p$ < .001, $\eta^2$ = .07. The mean rank sexual harassment score for gender diverse respondents was 1017.04, for women was 1053.89, and for men was 895.54. Dunn post-hoc tests with Bonferroni correction indicated that the mean rank sexual harassment score for women was significantly higher than for men ($p$ < .001). The mean rank sexual harassment score for gender diverse respondents was significantly higher than for men ($p$ = .002*)*. However, there was no significant difference between women and gender diverse respondents ($p$ = .91).

### Experienced Sexual Assault

A Kruskal-Wallis test indicated that there was a significant difference in experienced sexual assault across the three groups,

H(2) = 26.31, $p < .001$, $\eta^2 = .01$. The mean rank sexual assault score for gender diverse respondents was 964.46, for women was 980.94, and for men was 945.48. Dunn post-hoc tests with Bonferroni correction indicated that the mean rank sexual assault score for women was significantly higher than for men ($p < .001$). Gender diverse respondents were not significantly different from men ($p = .97$) or women ($p = 1.00$).

### Awareness of Personal Harassment

A Kruskal-Wallis test indicated that there was a significant difference in awareness of personal harassment across the three groups, H(2) = 31.70, $p < .001$, $\eta^2 = .02$. The mean rank awareness of personal harassment score for gender diverse respondents was 1174.76, for women was 1016.68, and for men was 910.45. Dunn post-hoc tests with Bonferroni correction indicated that the mean rank awareness of personal harassment score for women was significantly higher than for men ($p < .001$). The mean rank awareness of personal harassment score for gender diverse respondents was significantly higher than for men ($p < .001$) and for women ($p = .04$).

### Awareness of Sexual Harassment

A Kruskal-Wallis test indicated that there was a significant difference in awareness of sexual harassment across the three groups, H(2) = 28.90, $p < .001$, $\eta^2 = .01$. The mean rank awareness of sexual harassment score for gender diverse respondents was 1165.01, for women was 993.94, and for men was 918.58. Dunn post-hoc tests with Bonferroni correction indicated that the mean rank awareness of sexual harassment score for women was significantly higher than for men ($p < .001$). The mean rank awareness of sexual harassment score for gender diverse respondents was significantly higher than for men ($p < .001$) and for women ($p = .006$).

### Awareness of Sexual Assault

A Kruskal-Wallis test indicated that there was a significant difference in awareness of sexual assault across the three groups, H(2) = 19.85, $p < .001$, $\eta^2 = .01$. The mean rank awareness of sexual assault score for gender diverse respondents was 1107.31, for women was 956.73, and for men was 944.54. Dunn post-hoc tests with Bonferroni correction indicated that the mean rank awareness of sexual assault score for gender diverse respondents was significantly higher than for men ($p < .001$) and for women ($p < .001$). The mean rank awareness of sexual assault scores for men and women were not significantly different ($p = 1.00$).

## 2. How did frequencies of experienced harm and awareness of harm differ across binary gender identities and academic position?

As described above, we found that women report more experienced personal harassment, sexual harassment, sexual assault and awareness of personal and sexual harassment than men. We wanted to assess the pervasiveness of this finding, and whether gender differences occur across academic career stages. To do this, we compared scores between men and women for undergraduate students, graduate students, and various faculty positions. Mann-Whitney U tests were used to assess the differences between men and women at each academic position. See Table S1. Note that we do not include gender diverse respondents in these analyses due to very small sample sizes for some academic positions.

**Table S1**

*Mann-Whitney U Results Comparing Men and Women for Seven Academic Positions*

| BSc (N = 674) | | MSc (N = 203) | | PhD (N = 318) | | Post-Doc (N = 89) | | Assistant Prof. (N = 63) | | Associate Prof. (N = 60) | | Full Prof./Dean (N = 136) | |
|---|---|---|---|---|---|---|---|---|---|---|---|---|---|
| U | p | U | p | U | p | U | p | U | p | U | p | U | p |
| **Experienced Personal Harassment** | | | | | | | | | | | | | |

| 38359.0 | <.001 | 3956.0 | .007 | 8012.0 | <.001 | 478.0 | <.001 | 274.0 | .04 | 197.5 | .003 | 1060.5 | <.001 |
|---|---|---|---|---|---|---|---|---|---|---|---|---|---|
| **Awareness of Personal Harassment** | | | | | | | | | | | | | |
| 41125.0 | .002 | 4215.5 | .19 | 8676.5 | .001 | 666.0 | .12 | 199.5 | .002 | 233.0 | .05 | 1255.0 | .07 |
| **Experienced Sexual Harassment** | | | | | | | | | | | | | |
| 39973.5.0 | <.001 | 4249.0 | .02 | 8663.0 | <.001 | 671.5 | .004 | 320.0 | .09 | 220.0 | <.001 | 1289.0 | <.001 |
| **Awareness of Sexual Harassment** | | | | | | | | | | | | | |
| 42657.5 | .02 | 4641.5 | .96 | 9463.5 | .02 | 651.5 | .05 | 245.0 | .02 | 279.0 | .38 | 1350.5 | .19 |
| **Experienced Sexual Assault** | | | | | | | | | | | | | |
| 45120.0 | .02 | 4282.5 | .01 | 10948.0 | .22 | 763.0 | .17 | 378.0 | 1.00 | 320.0 | .13 | 1552.05 | .09 |
| **Awareness of Sexual Assault** | | | | | | | | | | | | | |
| 46036.0 | .95 | 4558.5 | .78 | 10985.5 | .95 | 788.5 | .63 | 287.0 | .002 | 325.0 | .59 | 1451.0 | .25 |

## Comparison Between Students and Professionals

We compared harm between students (N =1192) and professionals (N=768) using Mann-Whitney U tests. See Table S2.

**Table S2**

*Mann-Whitney U Tests Comparing Students and Professionals by Type of Tarm.*

| | U | p |
|---|---|---|
| **Experienced Personal Harassment** | 519270.6 | <.001 |
| **Awareness of Personal Harassment** | 543984.0 | <.001 |
| **Experienced Sexual Harassment** | 462753.0 | .048 |
| **Awareness of Sexual Harassment** | 489232.0 | <.001 |
| **Experienced Sexual Assault** | 451061.5 | .68 |
| **Awareness of Sexual Assault** | 455265.5 | .09 |

### 3.1 How did frequencies of experienced harm and awareness of harm differ across binary gender identities and race?

For this analysis, we focused on people who identified as either women or men and compared those who were White with those who were BIPOC (Black, Indigenous, or People of Colour). We did not include gender diverse respondents in this analysis given low sample sizes of gender diverse respondents when further categorized by race. Specifically, we compared BIPOC women, BIPOC men, White women, and White men. Based on intersectionality theory, we expected BIPOC women to report more experienced harm than BIPOC men, White women and White men, and White women and BIPOC men to report more experiences of harm than White men. We examined the difference between White women and BIOPC men in an exploratory fashion. Awareness of harm was analyzed in an exploratory fashion. Six KW tests with Dunn post-hoc tests with Bonferroni corrections were conducted.

#### *Experienced Personal Harassment*

A Kruskal-Wallis test indicated that there was a significant difference in experienced personal harassment across the four groups, $H(3) = 107.62$, $p < .001$, $\eta^2 = .06$. The mean rank personal harassment score for BIPOC women was 1083.26, for White women was 1031.92, for BIPOC men was 847.94, and for White men was 849.46. Dunn post-hoc tests with Bonferroni correction indicated that the mean rank personal harassment score for BIPOC women was significantly higher than for BIPOC men ($p < .001$) and White men ($p < .001$). The mean rank personal harassment score for White women was significantly higher than for BIPOC men ($p < .001$) and White men ($p < .001$). BIPOC women and White women were not significantly different ($p = .72$). BIPOC men and White men were not significantly different ($p = 1.00$).

### Experienced Sexual Harassment

A Kruskal-Wallis test indicated that there was a significant difference in experienced sexual harassment across the four groups, H(3) = 139.11, $p < .001$, $\eta^2 = .07$. The mean rank sexual harassment score for BIPOC women was 985.63, for White women was 1019.58, for BIPOC men was 858.36, and for White men was 859.37. Dunn post-hoc tests with Bonferroni correction indicated that the mean rank sexual harassment score for BIPOC women was significantly higher than for BIPOC men ($p < .001$) and White men ($p < .001$). The mean rank sexual harassment score for White women was higher than for BIPOC men ($p < .001$) and White men ($p < .001$). BIPOC women and White women were not significantly different ($p = .74$). BIPOC men and White men were not significantly different ($p = 1.00$).

### Experienced Sexual Assault

A Kruskal-Wallis test indicated that there was a significant difference in experienced sexual assault across the four groups, H(3) = 24.49, $p < .001$, $\eta^2 = .01$. The mean rank sexual assault score for BIPOC women was 945.58, for White women was 932.88, for BIPOC men was 901.12 and for White men was 907.91. Dunn post-hoc tests with Bonferroni correction indicated that the mean rank sexual assault score for BIPOC women was significantly higher than for BIPOC men ($p = .001$) and White men ($p = .002$). The mean rank sexual assault score for White women was higher than for BIPOC men ($p = .008$) and White men ($p$

= .006). BIPOC women and White women were not significantly different ($p$ = 1.00). BIPOC men and White men were not significantly different ($p$ = 1.00).

### *Awareness of Personal Harassment*

A Kruskal-Wallis test indicated that there was a significant difference in awareness of personal harassment across the four groups, H(3) = 29.22, $p$ <.001, $\eta^2$ = .01. The mean rank awareness of personal harassment score for BIPOC women was 979.59, for White women was 978.45, for BIPOC men was 797.38, and for White men was 908.43. Dunn post-hoc tests with Bonferroni correction indicated that the mean rank awareness of personal harassment score for BIPOC women was significantly higher than for BIPOC men ($p$ <.001) but not different from White men ($p$ = .35; unadjusted $p$ = .06). The mean rank awareness of personal harassment score for White women was significantly higher than for BIPOC men ($p$ < .001) but not different from White men ($p$ = .07; unadjusted $p$ = .01). BIPOC and White women were not significantly different ($p$ = 1.00). The mean rank awareness of personal harassment score for White men was higher than for BIPOC men ($p$ = .005).

### *Awareness of Sexual Harassment*

A Kruskal-Wallis test indicated that there was a significant difference in awareness of sexual harassment across the four groups, H(3) = 22.96, $p$ <.001, $\eta^2$ = .01. The mean rank awareness of sexual harassment score for BIPOC women was 906.99, for White women was 978.59, for BIPOC men was 833.98, and for White men was 904.78. Dunn post-hoc tests with Bonferroni correction indicated that the mean rank awareness of sexual harassment score for BIPOC women was not significantly different from BIPOC men ($p$ =.33, unadjusted $p$ = .06) nor White men ($p$ = 1.00). The mean rank awareness of sexual harassment score for White women was significantly higher than for BIPOC men ($p$ < .001) and White men ($p$ = .01). BIPOC and White women were not significantly different ($p$ = .23, unadjusted $p$ = .04). BIPOC men and White men were not significantly different ($p$ = .08, unadjusted $p$ = .01).

### *Awareness of Sexual Assault*

A Kruskal-Wallis test indicated that there were no significant differences in awareness of sexual assault across the four groups, H(3) = 1.84, $p$ = .61, $\eta^2 < .001$. The mean rank awareness of sexual assault score for BIPOC women was 923.37, for White women was 917.40 for BIPOC men was 894.60, and for White men was 916.41.

### 3.2 Comparing Black and Indigenous, POC, and White Respondents

We combined Black and Indigenous respondents due to small sample size and compared them with POC and White respondents.

### *Experienced Personal Harassment*

A Kruskal-Wallis test indicated that there was no significant difference in experienced personal harassment among Black and Indigenous women, POC women, and White women, H(2) = 1.80, $p$ =.41, $\eta^2 < .001$.

A Kruskal-Wallis test indicated that there was no significant difference in experienced personal harassment among Black and Indigenous men, POC men, and White men, H(2) = 4.6, $p$ = .10, $\eta^2$ = .002. The mean rank personal harassment score for Black and Indigenous men was 644.58, for POC men was 557.01, and for White men was 566.02. Dunn post-hoc tests were conducted, with Bonferroni correction. There was some evidence that Black and Indigenous men reported more personal harassment than POC men ($p$ = .10, unadjusted p = .03) and White men ($p$ = .14, unadjusted $p$ = .05). POC and White men were not different ($p$ = 1.00).

### *Experienced Sexual Harassment*

A Kruskal-Wallis test indicated that there was no significant difference in experienced sexual harassment among Black and Indigenous women, POC women, and White women, H(2) = 2.13, $p$ =.35, $\eta^2 < .001$.

A Kruskal-Wallis test indicated that there was no significant difference in experienced sexual harassment among Black and Indigenous men, POC men, and White men, H(2) = 5.30, $p$ = .07, $\eta^2$ = .003. The mean rank personal harassment score for Black and Indigenous men was 599.5, for POC men was 556.3, and for White men was 561.10. Dunn post-hoc tests were conducted, with Bonferroni correction. There was some evidence that Black and Indigenous men reported more sexual harassment than POC ($p$ = .06, unadjusted p = .02) and White men ($p$ = .10, unadjusted $p$ = .03). POC and White men were not different ($p$ = 1.00).

### Experienced Sexual Assault

A Kruskal-Wallis test indicated that there was no significant difference in experienced sexual assault among Black and Indigenous women, POC women, and White women, H(2) = .70, $p$ =.71, $\eta^2$ < .001.

A Kruskal-Wallis test indicated that there was no significant difference in experienced sexual assault among Black and Indigenous men, POC men, and White men, H(2) = 1.38, $p$ = .50, $\eta^2$ < .001.

### Awareness of Personal Harassment

A Kruskal-Wallis test indicated that there was no significant difference in awareness of personal harassment among Black and Indigenous women, POC women, and White women, H(2) = .83, $p$ =661, $\eta^2$ < .001.

A Kruskal-Wallis test indicated that there was a significant difference in awareness of personal harassment among Black and Indigenous men, POC men, and White men, H(2) = 17.85, $p$ < .001, $\eta^2$ = .01. The mean rank score for Black and Indigenous men was 632.7, for POC men was 495.8, and for White men was 579.7. Dunn post-hoc tests were conducted, with Bonferroni correction. Black and Indigenous men ($p$ = .05) and White men ($p$ <.001) reported more than POC men. White men and Black and Indigenous men were not different ($p$ = 1.00).

### Awareness of Sexual Harassment

A Kruskal-Wallis test indicated that there was no significant difference in awareness of sexual harassment among Black and Indigenous women, POC women, and White women, H(2) = 3.86, $p$ =.15, $\eta^2$ = .003.

A Kruskal-Wallis test indicated that there was a significant difference in awareness of sexual harassment among Black and Indigenous men, POC men, and White men, H(2) = 12.56, $p$ = .002, $\eta^2$ = .009. The mean rank score for Black and Indigenous men was 629.8, for POC men was 513.7, and for White men was 569.4. Dunn post-hoc tests were conducted, with Bonferroni correction. Black and Indigenous men ($p$ = .05) and White men ($p$ =.005) reported more than POC men. White men and Black and Indigenous men were not different ($p$ = .57).

### *Awareness of Sexual Assault*

A Kruskal-Wallis test indicated that there was no significant difference in awareness of sexual assault among Black and Indigenous women, POC women, and White women, H(2) = 1.08, $p$ =.58, $\eta^2$ < .001.

A Kruskal-Wallis test indicated that there was no significant difference in awareness of sexual assault across the four groups, H(3) = 1.82, $p$ = .61, $\eta^2$ < .001.

## 4. How did frequencies of experienced harm and awareness of harm differ across binary gender identities and disability?

We conducted six KW tests and Dunn post-hoc tests with Bonferroni corrections comparing disabled women, disabled men, women without disabilities, and men without disabilities. Although gender diverse respondents were excluded from this analysis due to small sample size, there was a higher percentage of gender diverse people with a disability (28.7%) than women with a disability (5.2%) and men with a disability (7.3%). Based on intersectionality theory, we expected disabled women to report more harm than women without disabilities, disabled men, and men without disabilities. We expected women without disabilities to report more harm than men without disabilities, and disabled men to report more harm than men without disabilities. The difference between women without disabilities and disabled men was examined in an exploratory fashion. Awareness of harm was analyzed in an exploratory fashion.

### Experienced Personal Harassment

A Kruskal-Wallis test indicated that there was a significant difference in experienced personal harassment across the four groups, H(3) = 119.37, $p$ <.001, $\eta^2$ = .07. The mean rank experienced personal harassment score for disabled women was 1083.65, for women without disabilities was 948.15, for disabled men was 873.36, and for men without disabilities was 765.16. Dunn post-hoc tests with Bonferroni correction indicated that the mean rank experienced personal harassment score for disabled women was significantly higher than for disabled men ($p$ =.008), women without disabilities ($p$ =.04), and men without disabilities ($p$ < .001). Women without disabilities reported more personal harassment than men without disabilities ($p$ < .001) but were not different from disabled men ($p$ = .73). Men with and without disabilities were not different from each other ($p$ = .13, uncorrected $p$ = .02).

### Experienced Sexual Harassment

A Kruskal-Wallis test indicated that there was a significant difference in experienced sexual harassment across the four groups, H(3) = 122.11, $p$ <.001, $\eta^2$ = .07. The mean rank experienced sexual harassment score for disabled women was 974.91, for women without disabilities was 910.82, for disabled men was 798.36, and for men without disabilities was 785.00. Dunn post-hoc tests with Bonferroni correction indicated that the mean rank experienced sexual harassment score for disabled women was significantly higher than for disabled men ($p$ < .001), and men without disabilities ($p$ < .001). Women without disabilities reported more sexual harassment than men without disabilities ($p$ < .001) and disabled men ($p$ = .002). Women with and without disabilities were not different from each other ($p$ =.29, unadjusted p-value, $p$ = .05). Men with and without disabilities were not different from each other ($p$ = 1.00).

### *Experienced Sexual Assault*

A Kruskal-Wallis test indicated that there was a significant difference in experienced sexual assault across the four groups, H(3) = 38.32, *p* <.001, η² = .02. The mean rank experienced sexual assault score for disabled women was 913.62, for women without disabilities was 848.91, for disabled men was 830.67, and for men without disabilities was 824.43. Dunn post-hoc tests with Bonferroni correction indicated that the mean rank experienced sexual assault score for disabled women was significantly higher than for disabled men (*p* = .001)*,* women without disabilities (*p* =.001), and men without disabilities (*p* < .001). Women without disabilities reported more sexual assault than men without disabilities (*p* = .001) but were not different from disabled men (*p* = 1.00). Men with and without disabilities were not different from each other (*p* = 1.00).

### *Awareness of Personal Harassment*

A Kruskal-Wallis test indicated that there was a significant difference in awareness of personal harassment across the four groups, H(3) = 25.61, *p* <.001, η² = .01. The mean rank awareness of personal harassment score for disabled women was 996.48, for women without disabilities was 881.72, for disabled men was 931.74, and for men without disabilities was 794.26. Dunn post-hoc tests with Bonferroni correction indicated that the mean rank awareness of personal harassment score for disabled women was significantly higher than for men without disabilities (*p* = .003), but not different from disabled men (*p* = 1.00) or women without disabilities (*p* = .33, unadjusted p-value *p* = .06). The mean rank awareness of personal harassment score for women without disabilities was significantly higher than men without disabilities (*p* = .001), but not different from disabled men (*p* = 1.00). Men with and without disabilities did not differ (*p* = .10, unadjusted *p* = .02).

### *Awareness of Sexual Harassment*

A Kruskal-Wallis test indicated that there was a significant difference in awareness of sexual harassment across the four groups, H(3) = 10.89, *p* =.01 η² = .005. The mean rank awareness of sexual harassment score for disabled women was 907.59, for women without disabilities was 866.53, for disabled men was 819.20, and for men without disabilities was 809.52. Dunn post-hoc tests with Bonferroni correction indicated that the mean rank awareness of sexual harassment score for disabled women was not significantly different than for disabled men ($p = 1.00$), nor from different from men without disabilities ($p = .29$, unadjusted p-value, $p = .05$), or women without disabilities ($p = 1.00$). The mean rank awareness of sexual harassment score for women without disabilities was significantly higher than men without disabilities ($p = .02$), but not different from disabled men ($p = 1.00$). Men with and without disabilities did not differ ($p = 1.00$).

*Awareness of Sexual Assault*

A Kruskal-Wallis test indicated that there was a no significant difference in awareness of sexual assault across the four groups, H(3) = 1.42, $p = .70$, $\eta^2 < .001$. The mean rank awareness of sexual assault score for disabled women was 854.45, for women without disabilities was 839.72, for disabled men was 832.64, and for men without disabilities was 827.57.

**5. How did frequencies of experienced harm and awareness of harm differ across binary gender identities and sexual orientation?**

Next, we conducted six KW tests and Dunns post-hoc tests with Bonferroni corrections comparing sexually diverse women, sexually diverse men, heterosexual women, and heterosexual men. Based on intersectionality theory, we expected sexually diverse women to report more harm than heterosexual women, sexually diverse men, and heterosexual men, and for sexually diverse men to report more harm than heterosexual men. We compared heterosexual women and sexually diverse men in an exploratory fashion. Awareness of harm was analyzed in an exploratory fashion.

*Experienced Personal Harassment*

A Kruskal-Wallis test indicated that there was a significant difference in experienced personal harassment across the four groups, H(3) = 109.25, $p < .001$, $\eta^2 = .06$. The mean rank experienced personal harassment score for sexually diverse women was 1033.04, for heterosexual women was 1046.36, for sexually diverse men was 825.17, and for heterosexual men was 843.05. Dunn post-hoc tests with Bonferroni correction indicated that the mean rank experienced personal harassment score for sexually diverse women was significantly higher than sexually diverse men ($p < .001$) and heterosexual men ($p < .001$). The mean rank experienced personal harassment score for heterosexual women was significantly higher than sexually diverse men ($p < .001$) and heterosexual men ($p < .001$). Sexually diverse and heterosexual women did not differ from each other ($p = 1.00$). Sexually diverse and heterosexual men did not differ from each other ($p = 1.00$).

### Experienced Sexual Harassment

A Kruskal-Wallis test indicated that there was a significant difference in experienced sexual harassment across the four groups, H(3) = 143.51, $p <.001$, $\eta^2 = .08$. The mean rank experienced sexual harassment score for sexually diverse women was 1040.35, for heterosexual women was 985.78, for sexually diverse men was 837.45, and for heterosexual men was 853.72. Dunn post-hoc tests with Bonferroni correction indicated that the mean rank experienced sexual harassment score for sexually diverse women was significantly higher than sexually diverse men ($p < .001$) and heterosexual men ($p < .001$). The mean rank experienced sexual harassment score for heterosexual women was significantly higher than sexually diverse men ($p < .001$) and heterosexual men ($p < .001$). Sexually diverse and heterosexual women did not differ from each other ($p = .07$, unadjusted p-value $p = .01$). Sexually diverse and heterosexual men did not differ from each other ($p = 1.00$).

### Experienced Sexual Assault

A Kruskal-Wallis test indicated that there was a significant difference in experienced sexual assault across the four groups, H(3) = 25.58, $p <.001$, $\eta^2 = .01$. The mean rank experienced sexual assault score for sexually diverse women was 933.71, for heterosexual women was 929.84, for sexually diverse men was 905.06, and for heterosexual men was 897.33. Dunn post-hoc tests with Bonferroni correction indicated that the mean rank experienced sexual assault score for sexually diverse women was significantly higher than for heterosexual men ($p$ = .002) but not sexually diverse men ($p$ = .38). The mean rank experienced sexual assault score for heterosexual women was significantly higher than heterosexual men ($p$ < .001) but not sexually diverse men ($p$ = .45). Sexually diverse and heterosexual women did not differ from each other ($p$ = 1.00). Sexually diverse and heterosexual men did not differ from each other ($p$ = 1.00).

### Awareness of Personal Harassment

A Kruskal-Wallis test indicated that there was a significant difference in awareness of personal harassment across the four groups, H(3) = 18.54, $p$ <.001, $\eta^2$ = .009. The mean rank awareness of personal harassment score for sexually diverse women was 989.56, for heterosexual women was 964.46, for sexually diverse men was 875.86, and for heterosexual men was 873.06. Dunn post-hoc tests with Bonferroni correction indicated that the mean rank awareness of personal harassment score for sexually diverse women was significantly higher than heterosexual men ($p$ = .006) but not sexually diverse men ($p$ = .22, unadjusted p-value $p$ = .04). The mean rank awareness of personal harassment score for heterosexual women was significantly higher than heterosexual men ($p$ = .004) but not sexually diverse men ($p$ = .44). Sexually diverse and heterosexual women did not differ from each other ($p$ = 1.00). Sexually diverse and heterosexual men did not differ from each other ($p$ = 1.00).

### Awareness of Sexual Harassment

A Kruskal-Wallis test indicated that there was a significant difference in awareness of sexual harassment across the four groups, H(3) = 13.203, $p$ <.001, $\eta^2$ = .006. The mean rank awareness of sexual harassment score for sexually diverse women was 972.80, for heterosexual women was 938.57, for sexually diverse men was 896.73, and for heterosexual men was 878.47. Dunn post-hoc tests with Bonferroni correction indicated that the mean rank awareness of sexual harassment score for sexually diverse women was significantly higher than heterosexual men ($p$ = .01), but not different from sexually diverse men ($p = .62$). Heterosexual women reported marginally statistically significantly more than heterosexual men ($p = .06$, unadjusted p value $p = .009$) but were not different from sexually diverse men ($p = 1.00$). Sexually diverse and heterosexual women did not differ ($p = 1.00$) and nor did sexually diverse and heterosexual men ($p = 1.00$).

### *Awareness of Sexual Assault*

A Kruskal-Wallis test indicated that there was no significant difference in awareness of sexual assault across the four groups, $H(3) = 3.83$, $p = .28$, $\eta^2 < .001$. The mean rank awareness of sexual assault score for sexually diverse women was 913.94, for heterosexual women was 911.81, for sexually diverse men was 946.06, and for heterosexual men was 898.32.

## Dalhousie University Sexualized Violence Policy and Personal Harassment Policy

Participants read the following prompt before answering the experiential questions: "When answering these questions please consider the following definitions, taken from Dalhousie University's Sexualized Violence Policy and Personal Harassment Policy."

"Personal Harassment" is abusive, unfair, or demeaning treatment of a person or group of persons that is known or ought reasonably to be known to be unwelcome and unwanted when: a. such treatment abuses the power one holds over another by virtue of their employment relationship or misuses authority associated with their position of employment; b. such treatment has the effect of seriously threatening or intimidating a person, and such treatment has the effect of unreasonably interfering with a person's or a group of person's employment or performance; c. such treatment has the effect of creating an intimidating, hostile, or offensive work environment.

"Sexual Harassment' refers to: a. vexatious sexual conduct or a course of comment that is known or ought reasonably to be known as unwelcome; b. a sexual solicitation or advance made to a person by another individual where the other individual is in a position to confer benefit on, or deny a benefit to, the person to whom the solicitation or advance is made, where the individual who makes the solicitation or advance knows or ought reasonably to know that it is unwelcome; or c. a reprisal or threat of reprisal against a person for rejecting a sexual solicitation or advance.

"Sexual Assault" refers to: a. any form of unwanted, forced, or coerced sexual activity, including kissing, fondling, touching, and any kind of intercourse, that is done onto someone without their consent; or b. any attempts or threats, by an act or a gesture, to force sexual activity onto someone, if the person committing the act had or caused person to believe the individual committing the act had the present ability to act on the attempt or threat.

# References

Ordak, M. (2023). Multiple comparisons and effect size: Statistical recommendations for authors planning to submit an article to        Allergy. *Allergy*, *78*(5). DOI: 10.1111/all.15700

Tomczak, M., & Tomczak, E. (2014). The need to report effect size estimates revisited. An overview of some recommended measures of effect size.